\documentclass[notitlepage,aps,proof]{revtex4-1}%
\usepackage{amssymb}
\usepackage{amsfonts}
\usepackage{amsmath}
\usepackage{graphicx}%
\providecommand{\U}[1]{\protect\rule{.1in}{.1in}}
\begin{document}
\title{Systematically extending classical nucleation theory}
\author{James F. Lutsko}
\affiliation{Center for Nonlinear Phenomena and Complex Systems CP 231, Universit\'{e}
Libre de Bruxelles, Blvd. du Triomphe, 1050 Brussels, Belgium}
\homepage{http://www.lutsko.com}
\email{jlutsko@ulb.ac.be}

\begin{abstract}
  The foundation for any discussion of first-order phse transitions is Classical Nucleation Theory(CNT). CNT, developed in the first half of the twentieth century, is based on a number of heuristically plausible assumtptions and the majority of theoretical work on nucleation is devoted to refining or extending these ideas. Ideally, one would like to {\em derive} CNT from a more fundamental description of nucleation so that its extension, development and refinement could be developed systematically.  In this paper, such a development is described based on a previously established (Lutsko, JCP 136:034509, 2012 ) connection between Classical Nucleation Theory and fluctuating hydrodynamics. Here, this connection is described without the need for artificial assumtions such as spherical symmetry.  The results are illustrated by application to CNT with moving clusters (a long-standing problem in the literature) and the  constructrion of CNT for ellipsoidal clusters. 
\end{abstract}
\date{\today }
\maketitle

\section{Introduction}

The process of the nucleation of first order phase transitions is of
importance across the range of scientific disciplines from chemistry and
physics to biology and materials science. From both the theoretical and the
experimental perspectives, its most challenging feature is that it is an
intrinsically multiscale problem. Small clusters of new phase forming in a
background of mother phase are thermodynamically unstable if they are below
the size of the critical cluster. They can only form and grow by a series of
thermal fluctuations and the formation of a critical cluster is consequently a
rare event. To observe nucleation under conditions of interest in many
applications requires macroscopic volumes of material and times scale as long
as hours or days even though the outcome - the critical cluster - is itself a
microscopic object with a size of micrometers or even nanometers. The fact
that the growing cluster, viewed as a subsystem of the total volume, is by
definition not in an equilibrium state makes the problem even more
challenging. As a result, nucleation remains an area of intense research by
experimentalists, theorists and simulators alike.

Despite - or, perhaps, because of - this complexity, the primary theoretical
description of nucleation has long been a collection of heuristic ideas known
collectively as Classical Nucleation Theory\ (CNT)\cite{Kashchiev}. The basic
idea of CNT is that clusters of new phase grow (or shrink)\ due to the
attachment (or detachment) of single monomeric growth units from (or to) the
mother phase. The rate of capture is calculated by treating the cluster as
being a quasi-static object that acts as a sink for mass and energy and then
calculating the rates of flow of mass and/or energy by whatever means is
appropriate to a given problem - e.g. by means of hydrodyamics for nucleation
in solution. To separate the rates of attachment and detachment of the
monomers for a cluster of $N$ growth units, a detailed balance condition is
invoked\cite{Kashchiev} to demand that the ratio be proportional to
$\exp\left(  -\beta\Delta F\left(  N,N+1\right)  \right)  $ where $\beta$ is
the inverse temperature and $\Delta F\left(  N,N+1\right)  =F\left(
N+1\right)  -F\left(  N\right)  $ is the free energy difference between
clusters of size $N$ and $N+1$. The free energy is approximated by a capillary
model:\ the cluster is assumed to be spherical with a sharp interface between
its interior and the mother phase. The free energy is the sum of a bulk term
scaling as the volume and a surface term scaling as the area of the cluster.

While CNT is the basis for a large part of the work on nucleation, it is
clearly a crude approximation and, indeed, has internal inconsistencies. For
example, in the capillary model for the free energy the density of the mother
phase outside the cluster is assumed to be constant but in the transport
calculation used to obtain the attachment rate, the density has a non-uniform
profile\cite{Kashchiev}. This and many other reasons have inspired a lot of
work aimed at improving CNT\cite{JFL_Chapter}. One target is the calculation
of the free energy of the cluster:\ it is relatively easy to improve the
capillary model by e.g. allowing the surface tension or the density inside the
cluster to depend on the cluster size\cite{Prestipino2012}. More generally,
classical Density\ Functional Theory\cite{Evans1979,lutsko:acp} provides very
accurate, microscopic models that can be used both to directly calculate the
properties and free energy of critical clusters as well as providing a basis
for the derivation of more coarse-grained descriptions up to and including the
capillary model\cite{lutsko:acp,Lutsko2011a}. In this sense, the problem of
the free energy may be considered to be solved.

The dynamics of CNT are more problematic. Ideally, one would like, in analogy
to the free energy, to have a more fundamental description from which CNT
could be derived. This would then presumeably allow for the systematic
improvement of the description in the same was as the DFT free energy serves
as a basis for more coarse-grained models. Recently, such a synthesis has been
proposed in which fluctuating hydrodynamics is used as a starting
point\cite{Lutsko_JCP_2011_Com,Lutsko_JCP_2012_1}. The DFT free energy is
introduced as a means of calculating the pressure thus addressing both the
free energy calculation and the dynamics at the same time and the resulting
theory termed Mesoscopic Nucleation Theory (MeNT). It has been shown that CNT
can be derived from this starting point by introducing appropriate
approximations and many other consequences of the theory have been developed.
Most particularly, it has been shown to give a rich and quite non-classical
description of nucleation, even for the simplest application of liquid-vapor
nucleation\cite{lutsko2014a}.

One problem with the work to date on MeNT has been that it relies heavily on
the assumption of spherical symmetry. As such, the main promise - of providing
a basis for nontrivial extensions of CNT - has so far been unfulfilled. The
goal of the present paper is to show how such extensions may be systematically
investigated. As in previous work on\ MeNT, attention will be restricted to
diffusion-limited nucleation in the over-damped limit which has the enormous
advantage that the full hydrodynamic description (involving density, momentum
and energy fields) reduces to a contracted descritpion formulated entirely in
terms of the density. Furthermore, diffusion-limited nucleation is of
considerable practical importance being applicable to the nucleation of
macromolecules in solution and to colloidal systems. As in the previous
development of MeNT, the key step for coarse graining will be the introduction
of parameterized density profiles. Based on the experience with
spherically-symmetric clusters, it is argued that the Fokker-Planck equation
must be covariant and that this fixes its structure once a metric is specified
for the space of coarse-graining parameters. The metric is, in turn, directly
determined from the full over-damped fluctuating hydrodynamics starting point
thus completing the theory. This theoretical development is the subject of
Section II of this paper. The third Section details applications to three
cases: first, the spherically symmetric results are re-derived using the new
approach, second a version of CNT is developed that allows for displacement of
the center of mass and finally the theory is developed for ellipsoidal
clusters haveing three independent degrees of freedom. The paper concludes
with perspectives for further developments.

\section{Theory}

\subsection{The dynamical model}

We take as the starting point the equation for the evolution of the local
density, $\rho_{t}\left(  \mathbf{r}\right)  $, as derived from fluctuating
hydrodynamics in the over-damped limit and using the DFT\ expression for the
pressure,%
\begin{equation}
\frac{\partial\rho_{t}\left(  \mathbf{r}\right)  }{\partial t}=D\nabla
\cdot\rho_{t}\left(  \mathbf{r}\right)  \nabla\frac{\delta\beta F\left[
\rho_{t}\right]  }{\delta\rho_{t}\left(  \mathbf{r}\right)  }+\nabla\cdot
\sqrt{2D\rho_{t}\left(  \mathbf{r}\right)  }\mathbf{\xi}_{t}\left(
\mathbf{r}\right)  \label{s}%
\end{equation}
where $D$ is the coefficient for tracer diffusion for the colloids, $F\left[
\rho_{t}\right]  $ is the so-called Helmholtz free energy functional coming
from DFT and $\mathbf{\xi}_{t}\left(  \mathbf{r}\right)  $ is the
(three-dimensional) white noise with correlation $\left\langle \xi_{t}%
^{a}\left(  \mathbf{r}\right)  \xi_{t^{\prime}}^{b}\left(  \mathbf{r}^{\prime
}\right)  \right\rangle =\delta^{ab}\delta\left(  t-t^{\prime}\right)
\delta\left(  \mathbf{r-r}^{\prime}\right)  $. A discussion of the derivation
of this equation from fluctuating hydrodynamics can be found in
Ref.(\cite{Lutsko_JCP_2012_1}). We take the point of view that this equation
as written is really a short-hand for a difference equation obtained by
discretizing in space and time. Using a standard discretization scheme based
on centered-differences, Eq.(\ref{s}) turns out to be Ito-Stratonovich
equivalent (ISE)\cite{Lutsko_JCP_2012_1}. The autocorrelation of the forces is
an operator,%
\begin{equation}
M_{tt^{\prime}}\left(  \mathbf{r},\mathbf{r}^{\prime}\right)  =\mathbf{\nabla
\cdot}\sqrt{2D\rho_{t}\left(  \mathbf{r}\right)  }\mathbf{\nabla}^{\prime
}\mathbf{\cdot}\sqrt{2D\rho_{t^{\prime}}\left(  \mathbf{r}^{\prime}\right)
}\delta\left(  t-t^{\prime}\right)  \delta\left(  \mathbf{r-r}^{\prime
}\right)
\end{equation}
where $\mathbf{\nabla}^{\prime}$ is the Laplacian for $\mathbf{r}^{\prime}$
and both Laplacians act on everything to their right. This can be simplified
by considering its action on a test function,%
\begin{align}
&  \int_{V}dr^{\prime}\int_{V}d\mathbf{r}^{\prime}M_{tt^{\prime}}\left(
\mathbf{r},\mathbf{r}^{\prime}\right)  f_{t^{\prime}}\left(  \mathbf{r}%
^{\prime}\right) \label{fde0}\\
&  =\int_{V}dr^{\prime}\int_{V}d\mathbf{r}^{\prime}\left[  \mathbf{\nabla
\cdot}\sqrt{2D\rho_{t}\left(  \mathbf{r}\right)  }\mathbf{\nabla}^{\prime
}\sqrt{2D\rho_{t^{\prime}}\left(  \mathbf{r}^{\prime}\right)  }\delta
_{tt^{\prime}}\delta\left(  \mathbf{r-r}^{\prime}\right)  \right]
f_{t^{\prime}}\left(  \mathbf{r}^{\prime}\right) \nonumber\\
&  =2D\mathbf{\nabla}\cdot\sqrt{\rho_{t}\left(  \mathbf{r}\right)  }\int%
_{V}\left[  \mathbf{\nabla}^{\prime}\sqrt{\rho_{t}\left(  \mathbf{r}^{\prime
}\right)  }\delta\left(  \mathbf{r-r}^{\prime}\right)  \right]  f_{t}\left(
\mathbf{r}^{\prime}\right)  d\mathbf{r}^{\prime}\nonumber\\
&  =-2D\mathbf{\nabla}\cdot\rho_{t}\left(  \mathbf{r}\right)  \mathbf{\nabla
}f_{t}\left(  \mathbf{r}\right)  +2D\left(  \mathbf{\nabla}\rho_{t}\left(
\mathbf{r}\right)  f_{t}\left(  \mathbf{r}\right)  \right)  \cdot\left(
\int_{\partial V}\delta\left(  \mathbf{r-r}^{\prime}\right)  d\mathbf{S}%
^{\prime}\right)  .\nonumber
\end{align}
where we assume that all functions of interest are nonzero only within a
volume $V$(the system volume) and we denote the surface of this volume as
$\partial V$. Assuming that the surface term can be neglected, the
autocorrelation is the same operator as acts on the free energy gradient in
the original SDE thus demonstrating the existence of a fluctuation-dissipation
relation (FDR) which in turn immediately implies that in equilibrium, the
probability to observe a given density configuration, $\rho\left(
\mathbf{r}\right)  $, is proportional to $\exp\left(  -\beta F\left[
\rho\right]  \right)  $ as one would expect.

\subsection{Transition probabilities and the geometry of density space}

The SDE describes the evolution of the density field $\rho_{t}\left(
\mathbf{r}\right)  $ and in this language, nucleation consists of a transition
from (the neighborhood of) an initial field $\rho^{\left(  i\right)  }\left(
\mathbf{r}\right)  $ describing the mother phase to a (the neighborhood of) a
final state $\rho^{\left(  f\right)  }\left(  \mathbf{r}\right)  $ describing
the new phase. In the simplest case of liquid-vapor nucleation, the initial
state would be a vapor for which the average density is a constant so
$\rho^{\left(  i\right)  }\left(  \mathbf{r}\right)  =\overline{\rho}_{v}$
where the vapor density $\overline{\rho}_{v}$is determined by the
thermodynamic conditions. The final state is a liquid for which the average
density is also constant so $\rho^{\left(  f\right)  }\left(  \mathbf{r}%
\right)  =\overline{\rho}_{l}$ with the liquid density $\overline{\rho}_{l}$
again being determined by the thermodynamics. More complicated states, such as
crystals, are of course also possible. One way to characterize the transition
is by specifying the nucleation pathway which is a sequence of density fields
starting with $\rho^{\left(  i\right)  }\left(  \mathbf{r}\right)  $ and
ending with $\rho^{\left(  f\right)  }\left(  \mathbf{r}\right)  $. The
sequence can be parameterized by some continuous index as $\rho_{\lambda
}\left(  \mathbf{r}\right)  $ for, say, $0\leq\lambda\leq1$ with $\rho
_{0}\left(  \mathbf{r}\right)  =\rho^{\left(  i\right)  }\left(
\mathbf{r}\right)  $ and $\rho_{1}\left(  \mathbf{r}\right)  =\rho^{\left(
f\right)  }\left(  \mathbf{r}\right)  $. Using generalizations of the
Onsager-Machlup formalism\cite{OnsagerMachlup}, the probability to make the
transition from the given initial state to the given final state can be
formulated as a path integral over the probability to observe any given
pathway with the latter being given by an expression of the form%
\begin{equation}
P\left(  \rho_{\lambda}\left(  \mathbf{r}\right)  \mid\rho_{0}\left(
\mathbf{r}\right)  ,\rho_{1}\left(  \mathbf{r}\right)  \right)  =\exp\left(
\int_{0}^{1}\mathcal{L}\left[  \rho_{\lambda}\left(  \mathbf{r}\right)
,\frac{d}{d\lambda}\rho_{\lambda}\left(  \mathbf{r}\right)  \right]
d\lambda\right)
\end{equation}
where explicit, exact expressions can be given for the Lagrangian functional
$\mathcal{L}$\cite{Graham}; in the weak noise limit, (for which the amplitude
of the noise is small compared to the deterministic term), the dominant
contribution is
\begin{equation}
\mathcal{L}\left[  \rho_{t}\right]  =\int_{V}\left(  \frac{\partial\rho
_{t}\left(  \mathbf{r}\right)  }{\partial t}-\mathbf{\nabla}\cdot\rho
_{t}\left(  \mathbf{r}\right)  \mathbf{\nabla}\frac{\delta\beta F\left[
\rho_{t}\right]  }{\delta\rho_{t}\left(  \mathbf{r}\right)  }\right)  \left(
\mathbf{\nabla}\cdot\rho_{t}\left(  \mathbf{r}\right)  \mathbf{\nabla}\right)
^{-1}\left(  \frac{\partial\rho_{t}\left(  \mathbf{r}\right)  }{\partial
t}-\mathbf{\nabla}\cdot\rho_{t}\left(  \mathbf{r}\right)  \mathbf{\nabla}%
\frac{\delta\beta F\left[  \rho_{t}\right]  }{\delta\rho_{t}\left(
\mathbf{r}\right)  }\right)  d\mathbf{r.} \label{LC}%
\end{equation}
This exact result allows one to ask for the \textit{most likely path} (MLP)
from the intial to the final state: namely, the path that maximizes the
transition probability which can be formulated as an Euler-Lagrange equation.

In the weak noise approximation, it is straightforward to demonstrate that the
FDR implies that the MLP\ passes through the critical cluster, defined as the
saddle point state $\rho^{\left(  c\right)  }\left(  \mathbf{r}\right)  $ for
which
\begin{equation}
\left(  \frac{\delta F\left[  \rho\right]  }{\delta\rho\left(  \mathbf{r}%
\right)  }\right)  _{\rho^{\left(  c\right)  }\left(  \mathbf{r}\right)  }=0,
\end{equation}
and that the energy barrier for nucleation is precisely $\Delta F=F\left[
\rho^{\left(  c\right)  }\right]  -F\left[  \rho^{\left(  i\right)  }\right]
$\cite{Lutsko_JCP_2012_1}. Thus, it would seem that the MLP\ is a good
candidate for a mathematically precise characterization of the "nucleation
pathway". In general, determining the MLP requires solving the Euler-Lagrange
equation which is second order in time, however in the case of barrier
crossing, there is a simpler alternative: the MLP\ can be constructed by
solving the first order equation%
\begin{equation}
\frac{\partial\rho_{\lambda}^{\left(  MLP\right)  }\left(  \mathbf{r}\right)
}{\partial\lambda}=D\mathbf{\nabla}\cdot\rho_{\lambda}^{\left(  MLP\right)
}\left(  \mathbf{r}\right)  \mathbf{\nabla}\frac{\delta\beta F\left[
\rho_{\lambda}^{\left(  MLP\right)  }\right]  }{\delta\rho_{\lambda}^{\left(
MLP\right)  }\left(  \mathbf{r}\right)  }%
\end{equation}
starting at the critical state, $\rho^{\left(  c\right)  }\left(
\mathbf{r}\right)  $, and perturbing infinitesimally in the direction of the
unstable (generalized) eigenvalue of the Hessian giving two paths:\ one
leading back to the initial state and one leading to the final state. The
union of these two paths is the MLP\ in the weak-noise
approximation\cite{Lutsko_JCP_2012_1}.

An important point in the present context is that the path is a geometric
object - time does not enter into the determination of the MLP. The
construction just described can be understood as gradient decent on the
potential energy surface $F\left[  \rho\right]  $ in density space with a
metric giving the distance between two infinitesimally close densities,
$\rho\left(  \mathbf{r}\right)  $ and $\rho\left(  \mathbf{r}\right)
+d\rho\left(  \mathbf{r}\right)  $, as
\begin{equation}
ds^{2}=\int_{V}\left\{  d\rho\left(  \mathbf{r}\right)  \left(
-\mathbf{\nabla}\cdot\rho\left(  \mathbf{r}\right)  \mathbf{\nabla}\right)
^{-1}d\rho\left(  \mathbf{r}\right)  \right\}  d\mathbf{r}%
\end{equation}
and the length of a path in density space is then
\begin{equation}
s=\int_{0}^{1}\sqrt{\int_{V}\left\{  \frac{d\rho_{\lambda}\left(
\mathbf{r}\right)  }{d\lambda}\left(  -\mathbf{\nabla}\cdot\rho_{\lambda
}\left(  \mathbf{r}\right)  \mathbf{\nabla}\right)  ^{-1}\frac{d\rho_{\lambda
}\left(  \mathbf{r}\right)  }{d\lambda}\right\}  d\mathbf{r}}d\lambda.
\end{equation}
The operator $\left(  \nabla\cdot\rho\left(  \mathbf{r}\right)  \nabla\right)
^{-1}$ is to be interpreted in the obvious way: for example the quantity
$\phi\left(  \mathbf{r}\right)  =\left(  \nabla\cdot\rho\left(  \mathbf{r}%
\right)  \nabla\right)  ^{-1}d\rho\left(  \mathbf{r}\right)  $ is determined
by solving
\begin{equation}
\mathbf{\nabla}\cdot\rho\left(  \mathbf{r}\right)  \mathbf{\nabla}\phi\left(
\mathbf{r}\right)  =d\rho\left(  \mathbf{r}\right)  \label{i1}%
\end{equation}
so that, e.g.,
\begin{equation}
ds^{2}=-\int_{V}d\rho\left(  \mathbf{r}\right)  \phi\left(  \mathbf{r}\right)
d\mathbf{r}%
\end{equation}
or, replaceing the factor of $d\rho\left(  \mathbf{r}\right)  $,
\begin{align}
ds^{2}  &  =-\int_{V}\left(  \mathbf{\nabla}\cdot\rho\left(  \mathbf{r}%
\right)  \mathbf{\nabla}\phi\left(  \mathbf{r}\right)  \right)  \phi\left(
\mathbf{r}\right)  d\mathbf{r}\\
&  =\int_{V}\rho\left(  \mathbf{r}\right)  \left(  \mathbf{\nabla}\phi\left(
\mathbf{r}\right)  \right)  ^{2}d\mathbf{r-}\int_{\partial V}\rho\left(
\mathbf{r}\right)  \phi\left(  \mathbf{r}\right)  \mathbf{\nabla}\phi\left(
\mathbf{r}\right)  \cdot d\mathbf{S.}\nonumber
\end{align}
where the second term on the right is a surface integral evaluated on the
boundaries enclosing the system. Since $ds^{2}$ must be non-negative and since
the first term on the right is obviously nonegative, we can ensure
non-negativity a sufficient condition is that either $\phi\left(
\mathbf{r}\right)  =0$ or $\nabla\phi\left(  \mathbf{r}\right)  =0$ on the surface.

In fact, the inverse of a differential operator only has meaning if the
corresponding boundary conditions are supplied:\ otherwise, there is no unique
potential $\phi\left(  \mathbf{r}\right)  $ making te problem ill-defined. The
boundary conditions follow from the physics of the original SDE, in the
present case Eq.(\ref{s}). This was derived using fluctuating hydrodynamics
and, in the case of a finite system, the most natural requirement is that the
total mass of the system be conserved. This means that the deviations in the
density, $d\rho\left(  \mathbf{r}\right)  $, must conserve the total mass,
i.e. that
\begin{equation}
dM=\int_{V}d\rho\left(  \mathbf{r}\right)  d\mathbf{r}=0.
\end{equation}
Integrating Eq.(\ref{i1}) then gives the no-flux condition%
\begin{equation}
\int_{\partial V}\rho\left(  \mathbf{r}\right)  \nabla\phi\left(
\mathbf{r}\right)  \cdot d\mathbf{S}=0.
\end{equation}
Further development depends on the details of the physical problem. In the
case of hard walls, we would restrict attention to the class of functions
satisfying the no-flux condition locally, $\nabla\phi\left(  \mathbf{r}%
\right)  \cdot d\mathbf{S=0}$ for all points on $\partial V$. This
automatically ensures that $ds^{2}$ is non-negative. It also completes the
argument for the fluctuation-dissipation relation by eliminating the last term
in Eq.(\ref{fde0}). For periodic boundaries, anything leaving via one wall
re-enters via another so that the no-flux condition is global but the
periodicity itself provides the boundary condition. Here, I will always
consider hard walls, conserved mass and the space of functions satisfying the
local no-flux condition.

Finally, we remark that the geometric interpretation given here is not
restricted to the weak-noise regime. In fact, using the language of
differential geometry, the strong-noise Lagrangian is fully covariant and the
MLP determined by it is independent of any choice of parameterization of
density space as discussed in classic papers by Graham\cite{Graham}.

\subsection{Order Parameters}

When determining the MLP, we must minize the transition probability with
respect to the density pathway. This can be done in an exact sense as
described above but we could also imagine a simpler, more restricted procedure
whereby we represent the density field by some parameterized form,
\begin{equation}
\rho\left(  \mathbf{r}\right)  =\rho\left(  \mathbf{r};\mathbf{x}\right)
\end{equation}
where $f$ is some fixed functional form that depends on a set of $N$ order
parameters, $x^{\alpha}$, for $\alpha=1,...,N$ . The distance between such a
density distribution and one with slightly different parameters, $\rho\left(
\mathbf{r};\mathbf{x}+d\mathbf{x}\right)  $, follows from the general
expression%
\begin{align}
ds^{2}  &  =\int_{V}\left\{  dx^{\alpha}\frac{\partial\rho\left(
\mathbf{r};\mathbf{x}\right)  }{\partial x^{\alpha}}\left(  -\mathbf{\nabla
}\cdot\rho\left(  \mathbf{r};\mathbf{x}\right)  \mathbf{\nabla}\right)
^{-1}\frac{\partial\rho\left(  \mathbf{r};\mathbf{x}\right)  }{\partial
x^{\beta}}dx^{\beta}\right\}  d\mathbf{r}\\
&  =g_{\alpha\beta}\left(  \mathbf{x}\right)  dx^{\alpha}dx^{\beta}\nonumber
\end{align}
where the metric in parameter space is
\begin{equation}
g_{\alpha\beta}\left(  \mathbf{x}\right)  =\int\left\{  \frac{\partial
\rho\left(  \mathbf{r};\mathbf{x}\right)  }{\partial x^{\alpha}}\left(
-\mathbf{\nabla}\cdot\rho\left(  \mathbf{r};\mathbf{x}\right)  \mathbf{\nabla
}\right)  ^{-1}\frac{\partial\rho\left(  \mathbf{r};\mathbf{x}\right)
}{\partial x^{\beta}}\right\}  d\mathbf{r.}%
\end{equation}
Defining $\phi_{\alpha}\left(  \mathbf{r;x}\right)  $ as%
\begin{equation}
\mathbf{\nabla}\cdot\rho\left(  \mathbf{r};\mathbf{x}\right)  \mathbf{\nabla
}\phi_{\alpha}\left(  \mathbf{r;x}\right)  =\frac{\partial\rho\left(
\mathbf{r};\mathbf{x}\right)  }{\partial x^{\alpha}}%
\end{equation}
this can also be written as%
\begin{align}
g_{\alpha\beta}\left(  \mathbf{x}\right)   &  =-\int_{V}\frac{\partial
\rho\left(  \mathbf{r};\mathbf{x}\right)  }{\partial x^{\alpha}}\phi_{\beta
}\left(  \mathbf{r;x}\right)  d\mathbf{r}\\
&  =\int_{V}\rho\left(  \mathbf{r};\mathbf{x}\right)  \left(  \mathbf{\nabla
}\phi_{\alpha}\left(  \mathbf{r;x}\right)  \right)  \cdot\left(
\mathbf{\nabla}\phi_{\beta}\left(  \mathbf{r;x}\right)  \right)
d\mathbf{r}-\int_{\partial V}\rho\left(  \mathbf{r};\mathbf{x}\right)
\phi_{\beta}\left(  \mathbf{r;x}\right)  \mathbf{\nabla}\phi_{\alpha}\left(
\mathbf{r;x}\right)  \cdot d\mathbf{S}\nonumber
\end{align}
The first term on the right displays the expected symmetry of the metric with
respect to the parameters: the vanishing of the surface term follows from the
no-flux condition.

Given the parameterization, the best approximation to the MLP would come from
minimizing the Lagrangian evaluated for such paths, namely (in the weak-noise
regime)
\begin{align}
L\left(  \mathbf{x}_{t}\right)   &  =\int_{V}\left(  \frac{\partial\rho\left(
\mathbf{r};\mathbf{x}_{t}\right)  }{\partial x_{t}^{\alpha}}\frac
{dx_{t}^{\alpha}}{dt}-\mathbf{\nabla}\cdot\rho\left(  \mathbf{r}%
;\mathbf{x}_{t}\right)  \mathbf{\nabla}\frac{\delta\beta F\left[  \rho
_{t}\right]  }{\delta\rho\left(  \mathbf{r};\mathbf{x}_{t}\right)  }\right)
\left(  -\mathbf{\nabla}\cdot\rho_{t}\left(  \mathbf{r}\right)  \mathbf{\nabla
}\right)  ^{-1}\\
&  \times\left(  \frac{\partial\rho\left(  \mathbf{r};\mathbf{x}_{t}\right)
}{\partial x_{t}^{\beta}}\frac{dx_{t}^{\beta}}{dt}-\mathbf{\nabla}\cdot
\rho\left(  \mathbf{r};\mathbf{x}_{t}\right)  \mathbf{\nabla}\frac{\delta\beta
F\left[  \rho_{t}\right]  }{\delta\rho\left(  \mathbf{r};\mathbf{x}%
_{t}\right)  }\right)  d\mathbf{r}\nonumber
\end{align}
which, upon expanding and using the definitions above and the functional chain
rule, can be written as
\begin{align}
L\left(  \mathbf{x}_{t}\right)   &  =\frac{dx_{t}^{\alpha}}{dt}g_{\alpha\beta
}\left(  \mathbf{x}_{t}\right)  \frac{dx_{t}^{\beta}}{dt}+2\frac
{dx_{t}^{\alpha}}{dt}\frac{\partial\beta F\left(  \mathbf{x}_{t}\right)
}{\partial x_{t}^{\alpha}}\\
&  +\int_{V}\frac{\delta\beta F\left[  \rho_{t}\right]  }{\delta\rho\left(
\mathbf{r};\mathbf{x}_{t}\right)  }\left(  -\mathbf{\nabla}\cdot\rho\left(
\mathbf{r};\mathbf{x}_{t}\right)  \mathbf{\nabla}\right)  \frac{\delta\beta
F\left[  \rho_{t}\right]  }{\delta\rho\left(  \mathbf{r};\mathbf{x}%
_{t}\right)  }d\mathbf{r}\nonumber
\end{align}

It is not possible to further simplify without more information. One case
amenable to analysis is that of a \emph{complete} parameterization. An example
would be to represent the density as an expansion in a complete set of basis
functions, $v_{i}\left(  \mathbf{r}\right)  $ , as
\begin{equation}
\rho_{t}\left(  \mathbf{r};\mathbf{x}\right)  =\sum_{i=1}^{\infty}x_{i}\left(
t\right)  v_{i}\left(  \mathbf{r}\right)  .
\end{equation}
Another would be, in the discretized version of the problem $\rho_{t}\left(
\mathbf{r}\right)  \rightarrow\rho_{i}\left(  t\right)  =\rho_{t}\left(
\mathbf{r}_{i}\right)  $, any invertible mapping $x_{i}=f_{i}\left(  \rho
_{0},..,\rho_{N}\right)  $. If the set of parameters is complete, then there
would have to be a completeness relations of the form
\begin{align}
\label{l1}\frac{\partial\rho\left(  \mathbf{r};\mathbf{x}\right)  }{\partial
x^{\alpha}}\frac{\delta x^{\alpha}}{\delta\rho\left(  \mathbf{r}^{\prime
};\mathbf{x}\right)  }  &  =\delta\left(  \mathbf{r-r}^{\prime}\right) \\
\int_{V}\frac{\delta x^{\alpha}}{\delta\rho\left(  \mathbf{r};\mathbf{x}%
\right)  }\frac{\partial\rho\left(  \mathbf{r};\mathbf{x}\right)  }{\partial
x^{\beta}}d\mathbf{r}  &  =\delta_{\beta}^{\alpha}\nonumber
\end{align}
and using this, one easily shows (see Appendix \ref{AppComplete} ) that the
weak-noise Lagrangian for the MLP with parmaterized paths is%
\begin{equation}
\mathcal{L}\left(  \mathbf{x}_{t}\right)  =\left(  \frac{dx_{t}^{\alpha}}%
{dt}+g^{\alpha\beta}\left(  \mathbf{x}_{t}\right)  \frac{\partial\beta
F\left(  \mathbf{x}_{t}\right)  }{\partial x_{t}^{\beta}}\right)
g_{\alpha\gamma}\left(  \mathbf{x}_{t}\right)  \left(  \frac{dx_{t}^{\gamma}%
}{dt}+g^{\gamma\sigma}\left(  \mathbf{x}_{t}\right)  \frac{\partial\beta
F\left(  \mathbf{x}_{t}\right)  }{\partial x_{t}^{\sigma}}\right)
\label{weak}%
\end{equation}
which has the same structure as the original continuum case, Eq.(\ref{LC}).

\subsection{Coarse-grained dynamics:\ the dynamics of parameterized paths}

Given that there is an induced metric in the space of order-parameters, it is
natural to ask if this can be viewed as arising directly from a stochastic
description so that one could speak of an order-parameter dynamics. If so, and
given the natural physical requirement of the existence of an equilibrium-like
state (at least for certain circumstances), one expects that any such
description would reduce in the weak-noise limit to
\begin{equation}
\frac{dx_{t}^{\alpha}}{dt}=-g^{\alpha\beta}\left(  \mathbf{x}_{t}\right)
\frac{\partial\beta F\left(  \mathbf{x}_{t}\right)  }{\partial x_{t}^{\beta}%
}+q_{a}^{\alpha}\left(  \mathbf{x}_{t}\right)  \xi_{t}^{a},\;\;\left\langle
\xi_{t}^{a}\xi_{t^{\prime}}^{a^{\prime}}\right\rangle =\delta^{aa^{\prime}%
}\delta\left(  t-t^{\prime}\right)  \label{sde1}%
\end{equation}
where the equilibrium state is ensured by an FDE, $g^{\alpha\beta}=\sum
_{a}q_{a}^{\alpha}q_{a}^{\beta}.$Note that the indices for the noise (written
using latin letters) are different in nature than the ones for the
order-parameters (greek letters):\ in fact, there is no requirement that the
number of noise terms be the same as the number of order-parameters. In
general, the existence of the FDE only demands that the number of noise terms
be sufficient that the matrix $q$ has enough degrees of freedom to satisfy the
FDE. Thus the noise-indices are simply indexes do not refer to geometric
coordinates. Such a weak-noise dynamics is also consistent with the
parameterized Lagrangian derived above.

The question we pose here is whether any further justification can be given
for this equation and if, so, can we say anything about the strong-noise
regime? There is probably no unique answer except in the special case of a
complete parameterization. In this case, a straightforward derivation, given
in Appendix \ref{AppDeriv} results in the Ito SDE%
\begin{align}
\frac{dx_{t}^{\alpha}}{dt}  & =-Dg^{\alpha\beta}\left(  \mathbf{x}_{t}\right)
\frac{\partial\beta F\left(  \mathbf{x}_{t}\right)  }{\partial x_{t}^{\beta}%
}+\frac{1}{2}Dg^{\alpha\beta}\left(  \mathbf{x}_{t}\right)  \frac{1}%
{\det\mathbf{g}\left(  \mathbf{x}_{t}\right)  }\frac{\partial}{\partial
x_{t}^{\beta}}\det\mathbf{g}\left(  \mathbf{x}_{t}\right)  \\
& -\frac{1}{2}DA^{\alpha}\left(  \mathbf{x}_{t}\right)  +D\frac{\partial
g^{\alpha\beta}\left(  \mathbf{x}\right)  }{\partial x^{\beta}}+\sqrt{2D}%
q_{a}^{\alpha}\left(  \mathbf{x}_{t}\right)  \mathbf{\xi}_{t}^{a}.\nonumber
\end{align}
where $q_{a}^{\alpha}\left(  \mathbf{x}_{t}\right)  q_{a}^{\beta}\left(
\mathbf{x}_{t}\right)  =g^{\alpha\beta}\left(  \mathbf{x}_{t}\right)  $ and
the new contribution to the deterministic driving force is
\begin{align}
A^{\alpha}\left(  \mathbf{x}\right)   &  =\left(  g^{\alpha\beta}\left(
\mathbf{x}\right)  g^{\gamma\sigma}\left(  \mathbf{x}\right)  -g^{\alpha
\gamma}\left(  \mathbf{x}\right)  g^{\beta\sigma}\left(  \mathbf{x}\right)
\right)  \\
&  \times\int d\mathbf{r\;}\frac{\partial\rho\left(  \mathbf{r;x}\right)
}{\partial x^{\beta}}\left(  \mathbf{\nabla}\left[  \left(  \mathbf{\nabla
}\cdot\rho\left(  \mathbf{r;x}\right)  \mathbf{\nabla}\right)  ^{-1}%
\frac{\partial\rho\left(  \mathbf{r;}x\right)  }{\partial x^{\gamma}}\right]
\right)  \cdot\left(  \mathbf{\nabla}\left[  \left(  \mathbf{\nabla}\cdot
\rho\left(  \mathbf{r;x}\right)  \mathbf{\nabla}\right)  ^{-1}\frac
{\partial\rho\left(  \mathbf{r;}x\right)  }{\partial x^{\sigma}}\right]
\right)  ,\nonumber
\end{align}
while the corresponding Fokker-Plank equation is%
\begin{equation}
\frac{\partial}{\partial t}P_{t}\left(  \mathbf{x}\right)  =\frac{\partial
}{\partial x^{\alpha}}\left(
\begin{array}
[c]{c}%
g^{\alpha\beta}\left(  \mathbf{x}\right)  D\frac{\partial\beta F\left(
\mathbf{x}\right)  }{\partial x^{\beta}}-\frac{1}{2}Dg^{\alpha\beta}\left(
\mathbf{x}\right)  \frac{1}{\det\mathbf{g}\left(  \mathbf{x}\right)  }%
\frac{\partial}{\partial x^{\beta}}\det\mathbf{g}\left(  \mathbf{x}\right)  \\
+\frac{1}{2}DA^{\alpha}\left(  \mathbf{x}\right)  -D\frac{\partial
g^{\alpha\beta}\left(  \mathbf{x}\right)  }{\partial x^{\beta}}+D\frac
{\partial}{\partial x^{\beta}}g^{\alpha\beta}\left(  \mathbf{x}\right)
\end{array}
\right)  P_{t}\left(  \mathbf{x}\right)  .\label{FPE}%
\end{equation}
The new contribution to the driving force, $A^{\alpha}\left(  \mathbf{x}%
_{t}\right)  $, vanishes in the case of a single variable and it is tempting
to neglect it in general. For example, without it, the equilibrium
distribution is easily seen to be $P_{eq}\left(  \mathbf{x}\right)  =N\left(
\det\mathbf{g}\left(  \mathbf{x}_{t}\right)  \right)  ^{1/2}\exp\left(  -\beta
F\mathbf{x}\right)  $ (where $N$ is a normalization constant) whereas with it,
we cannot explicitly determine the equilibrium distribution. Fortunately, in
the following, we will mostly be concerned with the weak noise limit, for
which analytic results are possible, and in which this term does not appear
and these equations reduce to the simple SDE proposed intuitively in
Eq.(\ref{sde1}).

\subsection{Spherically Symmetric systems}

If all quantities are spherically symmetric and if the system volume is a
sphere of radius $R_{V}$, then it is straitforward to show that the inverse
operator needed for the metric and consistent with the boundary conditions is
\begin{equation}
\left(  -\nabla\cdot\rho\left(  r;\mathbf{x}_{t}\right)  \nabla\right)
^{-1}f\left(  r\right)  =A+\int_{0}^{r}\frac{1}{4\pi r^{\prime2}\rho\left(
r^{\prime};\mathbf{x}\right)  }\left(  \int\Theta\left(  r^{\prime}%
-r^{\prime\prime}\right)  f\left(  r^{\prime\prime};\mathbf{x}\right)
d\mathbf{r}^{\prime\prime}\right)  dr^{\prime}%
\end{equation}
where the integration constant $A$ is arbitrary. It is then easy to evaluated
the metric with the result%
\begin{equation}
g_{\alpha\beta}\left(  \mathbf{x}\right)  =-\int_{V}\left\{  \frac
{\partial\rho\left(  r;\mathbf{x}\right)  }{\partial x^{\alpha}}\int_{0}%
^{r}\frac{1}{4\pi r^{\prime2}\rho\left(  r^{\prime};\mathbf{x}\right)  }%
\frac{\partial m\left(  r^{\prime};\mathbf{x}\right)  }{\partial x^{\beta}%
}dr^{\prime}\right\}  d\mathbf{r}%
\end{equation}
where the cumulative mass up to radius $r$ is%
\begin{equation}
m\left(  r;\mathbf{x}\right)  =\int\Theta\left(  r-r^{\prime}\right)
\rho\left(  r^{\prime};\mathbf{x}\right)  d\mathbf{r}^{\prime}=4\pi\int%
_{0}^{r}r^{\prime2}\rho\left(  r^{\prime};\mathbf{x}\right)  dr^{\prime}.
\end{equation}
Integration by parts gives%
\begin{equation}
g_{\alpha\beta}\left(  \mathbf{x}\right)  =\int_{0}^{R_{V}}\frac{1}{4\pi
r^{2}\rho\left(  r;\mathbf{x}\right)  }\frac{\partial m\left(  r;\mathbf{x}%
\right)  }{\partial x^{\alpha}}\frac{\partial m\left(  r;\mathbf{x}\right)
}{\partial x^{\beta}}dr. \label{gsph}%
\end{equation}
The anomolous flux is calculated using%
\begin{align}
\nabla\left[  \left(  \nabla\cdot\rho\left(  r\mathbf{;x}\right)
\nabla\right)  ^{-1}\frac{\partial\rho\left(  r\mathbf{;x}\right)  }{\partial
x^{\sigma}}\right]   &  =\nabla\left[  A+\int_{0}^{r}\frac{1}{4\pi r^{\prime
2}\rho\left(  r^{\prime};\mathbf{x}\right)  }\frac{\partial m\left(
r^{\prime}\mathbf{;x}\right)  }{\partial x^{\sigma}}dr^{\prime}\right] \\
&  =\widehat{\mathbf{r}}\frac{1}{4\pi r^{2}\rho\left(  r;\mathbf{x}\right)
}\frac{\partial m\left(  r\mathbf{;x}\right)  }{\partial x^{\sigma}}\nonumber
\end{align}
giving%
\begin{equation}
A^{\alpha}\left(  \mathbf{x}\right)  =\left(  g^{\alpha\beta}\left(
\mathbf{x}\right)  g^{\gamma\sigma}\left(  \mathbf{x}\right)  -g^{\alpha
\gamma}\left(  \mathbf{x}\right)  g^{\beta\sigma}\left(  \mathbf{x}\right)
\right)  \int_{0}^{R_{\infty}}\mathbf{\;}\frac{1}{4\pi r^{2}\rho^{2}\left(
r;\mathbf{x}\right)  }\frac{\partial\rho\left(  r\mathbf{;x}\right)
}{\partial x^{\beta}}\frac{\partial m\left(  r\mathbf{;x}\right)  }{\partial
x^{\gamma}}\frac{\partial m\left(  r\mathbf{;x}\right)  }{\partial x^{\sigma}%
}dr.
\end{equation}
These results reproduce those previously derived using spherical symmetry from
the beginning\cite{Lutsko_JCP_2011_Com,Lutsko_JCP_2012_1}.

\subsection{Nucleation rates}

The most important question from a practical point of view is the nucleation
rate. This is related to the mean first passage time for barrier crossing. For
the one dimensional case, there is an exact expression for this quantity,
however, in the general case no such result exists. The standard
result\cite{Langer1,Talkner,Hanggi,Gardiner} valid in the weak noise limit is,
in our language,%
\begin{equation}
t_{mfp}=\varepsilon^{-1}\frac{\pi}{D\left\vert \lambda_{-}\right\vert }%
\frac{\sqrt{\left\vert \det\beta F_{\alpha\beta}^{\left(  c\right)
}\right\vert }}{\sqrt{\det\mathbf{g}(\mathbf{x}_{c})}\left(  2\pi\right)
^{N/2}}\left(  \int_{\Sigma_{meta}}\sqrt{\det\mathbf{g}\left(  \mathbf{x}%
\right)  }e^{-\beta F\left(  \mathbf{x}\right)  }d\mathbf{x}\right)  e^{\beta
F(\mathbf{x}_{c})} \label{T}%
\end{equation}
where $F_{\alpha\beta}^{\left(  c\right)  }$ is the Hessian of the free energy
evaluated at the critical cluster $\mathbf{x}_{c}$, $\lambda_{-}$ the (sole)
negative eigenvalue of $g^{\alpha\beta}(\mathbf{x}_{c})F_{\beta\gamma
}^{\left(  c\right)  }$ and $N$ is the number of order parameters. The
critical cluster is determined as usual by $\left.  \partial F(\mathbf{x}%
)/\partial x^{\alpha}\right\vert _{\mathbf{x}_{c}}=0$. The integral on the
right is a measure of the occupation of the metastable basin and the domain of
integration is restricted to this region.

\section{Classical Nucleation Theory and Generalizations}

\subsection{Generalized Classical Nucleation Theory}

\subsubsection{The density in CNT}

We take "CNT" to be the following elements:\ a sharp interface between the
cluster and bath, uniform density inside and outside the cluster and the
capillary model for the free energy. Mathematically, the sharp interface means
that there is an indicator function, $\chi\left(  \mathbf{r}\right)  $, which
is equal to one inside the cluster and zero outside so that, with the second
element of constant densities, we have that%
\begin{equation}
\rho\left(  \mathbf{r};\mathbf{x}\right)  =\rho_{0}\left(  \mathbf{x}\right)
\chi\left(  \mathbf{r;x}\right)  +\rho_{\infty}\left(  \mathbf{x}\right)
\left(  1-\chi\left(  \mathbf{r;x}\right)  \right)  .
\end{equation}

\subsubsection{The CNT\ metric}

The equation to be solved for the potential is%
\begin{equation}
\mathbf{\nabla}\rho\left(  \mathbf{r};\mathbf{x}\right)  \mathbf{\nabla}%
\phi_{\alpha}\left(  \mathbf{r};\mathbf{x}\right)  =\left(  \rho_{0}\left(
\mathbf{x}\right)  -\rho_{\infty}\left(  \mathbf{x}\right)  \right)
\frac{\partial\chi\left(  \mathbf{r;x}\right)  }{\partial x^{\alpha}}%
+\frac{\partial\rho_{0}\left(  \mathbf{x}\right)  }{\partial x^{\alpha}}%
\chi\left(  \mathbf{r;x}\right)  +\frac{\partial\rho_{\infty}\left(
\mathbf{x}\right)  }{\partial x^{\alpha}}\left(  1-\chi\left(  \mathbf{r;x}%
\right)  \right)
\end{equation}
so that we postulate a solution of the form
\begin{equation}
\phi_{\alpha}\left(  \mathbf{r};\mathbf{x}\right)  =\phi_{\alpha}^{\left(
0\right)  }\left(  \mathbf{r};\mathbf{x}\right)  \chi\left(  \mathbf{r;x}%
\right)  +\phi_{\alpha}^{\left(  1\right)  }\left(  \mathbf{r};\mathbf{x}%
\right)  \left(  1-\chi\left(  \mathbf{r;x}\right)  \right)  .
\end{equation}
Now, in general, because of its discontinuous nature, the Laplacian of the
indicator function will be proportional to a Dirac delta function. In fact, if
the surface is described by an equation of the form $\psi\left(
\mathbf{r;x}\right)  =0$ and the interior of the cluster by $\psi\left(
\mathbf{r;x}\right)  >0$ then%
\begin{equation}
\chi\left(  \mathbf{r;x}\right)  =\Theta\left(  \psi\left(  \mathbf{r;x}%
\right)  \right)
\end{equation}
and%
\begin{equation}
\mathbf{\nabla}\chi\left(  \mathbf{r;x}\right)  =\left(  \mathbf{\nabla}%
\psi\left(  \mathbf{r;x}\right)  \right)  \delta\left(  \psi\left(
\mathbf{r;x}\right)  \right)  .
\end{equation}
As a trivial example, for a sphere of radius $R$ we have $\psi\left(
\mathbf{r;x}\right)  =R-r$. Hence, substituting the ansatz for the potential
into the Poisson equation and equating coefficients of the delta function and
its derivatives gives
\begin{align}
\rho_{0}\nabla^{2}\phi_{\alpha}^{\left(  0\right)  }  &  =\frac{\partial
\rho_{0}}{\partial x^{\alpha}},\;\chi\left(  \mathbf{r}\right)  =1\\
\rho_{1}\nabla^{2}\phi_{\alpha}^{\left(  1\right)  }  &  =\frac{\partial
\rho_{1}}{\partial x^{\alpha}},\;\chi\left(  \mathbf{r}\right)  =0\nonumber
\end{align}
with the boundary conditions on the surface of the cluster%
\begin{align}
\phi_{\alpha}^{\left(  0\right)  }  &  =\phi_{\alpha}^{\left(  1\right)
},\;\psi\left(  \mathbf{r;x}\right)  =0\\
\left(  \rho_{0}\mathbf{\nabla}\phi_{\alpha}^{\left(  0\right)  }-\rho
_{1}\mathbf{\nabla}\phi_{\alpha}^{\left(  1\right)  }\right)  \cdot
\mathbf{\nabla}\psi\left(  \mathbf{r;x}\right)   &  =\left(  \rho_{0}-\rho
_{1}\right)  \frac{\partial\psi\left(  \mathbf{r;x}\right)  }{\partial
x^{\alpha}},\;\psi\left(  \mathbf{r;x}\right)  =0\nonumber
\end{align}
and the global no-flux boundary condition.

\subsubsection{The CNT free energy and critical cluster}

Finally, the capillary model generalizes to%
\begin{equation}
F\left(  \mathbf{x}\right)  =f\left(  \rho_{0}\left(  \mathbf{x}\right)
\right)  V\left(  \mathbf{x}\right)  +f\left(  \rho_{1}\left(  \mathbf{x}%
\right)  \right)  \left(  V-V\left(  \mathbf{x}\right)  \right)
+\gamma\left(  \mathbf{x}\right)  S\left(  \mathbf{x}\right)
\end{equation}
where $S\left(  \mathbf{x}\right)  $ and $V\left(  \mathbf{x}\right)  $ are
the surface and volume of the cluster respectively, $\gamma$ is the surface
tension between the two phases and $f\left(  \rho\right)  $ is the Helmholtz
free energy of the homogeneous bulk system. In general, the initial, final and
critical states will satisfy%
\begin{align}
0  &  =V\left(  \mathbf{x}\right)  \frac{\partial}{\partial x^{\alpha}%
}f\left(  \rho_{0}\left(  \mathbf{x}\right)  \right)  +\left(  V-V\left(
\mathbf{x}\right)  \right)  \frac{\partial}{\partial x^{\alpha}}f\left(
\rho_{1}\left(  \mathbf{x}\right)  \right)  +\left(  f\left(  \rho_{0}\left(
\mathbf{x}\right)  \right)  -f\left(  \rho_{1}\left(  \mathbf{x}\right)
\right)  \right)  \frac{\partial V\left(  \mathbf{x}\right)  }{\partial
x^{\alpha}}\\
&  +\gamma\left(  \mathbf{x}\right)  \frac{\partial S\left(  \mathbf{x}%
\right)  }{\partial x^{\alpha}}+\frac{\partial\gamma\left(  \mathbf{x}\right)
}{\partial x^{\alpha}}S\left(  \mathbf{x}\right) \nonumber
\end{align}
Since we are restricting attention to profiles that conserve mass, there is a
constraint%
\begin{equation}
M=\rho_{0}\left(  \mathbf{x}\right)  V\left(  \mathbf{x}\right)  +\left(
V-V\left(  \mathbf{x}\right)  \right)  \rho_{1}\left(  \mathbf{x}\right)
\end{equation}
giving%
\begin{equation}
0=\frac{\partial\rho_{0}\left(  \mathbf{x}\right)  }{\partial x^{\alpha}%
}V\left(  \mathbf{x}\right)  +\frac{\partial\rho_{1}\left(  \mathbf{x}\right)
}{\partial x^{\alpha}}\left(  V-V\left(  \mathbf{x}\right)  \right)  +\left(
\rho_{0}\left(  \mathbf{x}\right)  -\rho_{0}\left(  \mathbf{x}\right)
\right)  \frac{\partial V\left(  \mathbf{x}\right)  }{\partial x^{\alpha}}%
\end{equation}
and%
\begin{align}
0  &  =V\left(  \mathbf{x}\right)  \left(  f^{\prime}\left(  \rho_{0}\left(
\mathbf{x}\right)  \right)  -f^{\prime}\left(  \rho_{1}\left(  \mathbf{x}%
\right)  \right)  \right)  \frac{\partial\rho_{0}\left(  \mathbf{x}\right)
}{\partial x^{\alpha}}\label{Stationary}\\
&  +\left(  \left\{  f\left(  \rho_{0}\left(  \mathbf{x}\right)  \right)
-\rho_{0}\left(  \mathbf{x}\right)  f^{\prime}\left(  \rho_{1}\left(
\mathbf{x}\right)  \right)  \right\}  -\left\{  f\left(  \rho_{1}\left(
\mathbf{x}\right)  \right)  -f^{\prime}\left(  \rho_{1}\left(  \mathbf{x}%
\right)  \right)  \rho_{1}\left(  \mathbf{x}\right)  \right\}  \right)
\frac{\partial V\left(  \mathbf{x}\right)  }{\partial x^{\alpha}}\nonumber\\
&  +\gamma\left(  \mathbf{x}\right)  \frac{\partial S\left(  \mathbf{x}%
\right)  }{\partial x^{\alpha}}+\frac{\partial\gamma\left(  \mathbf{x}\right)
}{\partial x^{\alpha}}S\left(  \mathbf{x}\right)  .\nonumber
\end{align}

\subsection{Classical Nucleation Theory}

We recover CNT by (a) demanding spherical symmetry; (b) taking the only
parameter to be the radius of the cluster, $R$; (c) assuming the surface
tension is independent of the radius; and (d) assuming the cluster radius is
small compared to the system size. With these approximations, all quantities
depend only on the cluster radius, $R$, and if the system is confined to a
spherical volume with total radius $R_{V}$, then the exterior density is
\begin{align}
\rho_{1}\left(  R\right)   &  =\frac{M-\rho_{0}V\left(  R\right)  }{V\left(
R_{V}\right)  -V\left(  R\right)  }\\
&  =\overline{\rho}+\left(  \overline{\rho}-\rho_{0}\right)  \left(  \frac
{R}{R_{V}}\right)  ^{3}+O\left(  \frac{R}{R_{V}}\right)  ^{6}\nonumber
\end{align}
where $M$ is the total mass and $\overline{\rho}=M/V\left(  R_{V}\right)  $ is
the average density. We calculate the metric directly from the closed
expression, Eq.(\ref{gsph}) and find that
\begin{equation}
g_{RR}=4\pi\frac{\left(  \rho_{0}-\overline{\rho}\right)  ^{2}}{\overline
{\rho}}R^{3}\left(  1+O\left(  \frac{R}{R_{V}}\right)  \right)  .
\end{equation}
The interior density, $\rho_{0}$, is, in\ CNT, taken to be the bulk
equilibrium density and therefore independent of the cluster radius. The
stationarity condition, Eq.(\ref{Stationary}) then determines the critical
radius
\begin{equation}
R_{C}=-\frac{2\gamma}{\left\{  f\left(  \rho_{0}\right)  -\rho_{0}f^{\prime
}\left(  \overline{\rho}\right)  \right\}  -\left\{  f\left(  \overline{\rho
}\right)  -f^{\prime}\left(  \overline{\rho}\right)  \overline{\rho}\right\}
}+O\left(  \frac{R}{R_{V}}\right)  ^{3}%
\end{equation}
and the corresponding barrier is
\begin{align}
\Delta\beta F_{c} &  =\beta F\left(  R_{c}\right)  -\beta F\left(  0\right)
\\
&  =\beta\left(  f\left(  \rho_{0}\right)  -\rho_{0}f^{\prime}\left(
\overline{\rho}\right)  -f\left(  \overline{\rho}\right)  +\overline{\rho
}f^{\prime}\left(  \overline{\rho}\right)  \right)  V\left(  R_{C}\right)
+\beta\gamma S\left(  R_{C}\right)  +O\left(  \frac{R}{R_{V}}\right)
^{3}\nonumber\\
&  =\frac{16}{3}\pi\frac{\beta\gamma^{3}}{\left(  \left\{  f\left(  \rho
_{0}\right)  -\rho_{0}f^{\prime}\left(  \overline{\rho}\right)  \right\}
-\left\{  f\left(  \overline{\rho}\right)  -f^{\prime}\left(  \overline{\rho
}\right)  \overline{\rho}\right\}  \right)  ^{2}}\nonumber
\end{align}
Note that the ambient gas plays the role of a resevoir fixing the chemical
potential at $\mu=f^{\prime}\left(  \overline{\rho}\right)  $. For arbitrary
radii, the excess free energy can be written in the compact form
\begin{equation}
\beta\Delta F\left(  R\right)  =\beta\Delta F_{c}\left(  2\left(  \frac
{R}{R_{c}}\right)  ^{3}-3\left(  \frac{R}{R_{c}}\right)  ^{2}\right)
+O\left(  \frac{R}{R_{V}}\right)  ^{3}.
\end{equation}
We can do something similar for the undersaturated fluid and in general, if we
define
\begin{align}
\widetilde{R} &  =\left\vert \frac{2\gamma}{\left\{  f\left(  \rho_{0}\right)
-\rho_{0}f^{\prime}\left(  \overline{\rho}\right)  \right\}  -\left\{
f\left(  \overline{\rho}\right)  -f^{\prime}\left(  \overline{\rho}\right)
\overline{\rho}\right\}  }\right\vert \\
\Delta\beta\widetilde{F} &  =\frac{16}{3}\pi\left\vert \frac{\beta\gamma^{3}%
}{\left(  \left\{  f\left(  \rho_{0}\right)  -\rho_{0}f^{\prime}\left(
\overline{\rho}\right)  \right\}  -\left\{  f\left(  \overline{\rho}\right)
-f^{\prime}\left(  \overline{\rho}\right)  \overline{\rho}\right\}  \right)
^{2}}\right\vert \nonumber
\end{align}
then
\begin{equation}
\beta\Delta F\left(  R\right)  =\Delta\beta\widetilde{F}\left[  \mp2\left(
\frac{R}{\widetilde{R}}\right)  ^{3}+3\left(  \frac{R}{\widetilde{R}}\right)
^{2}\right]  +O\left(  \frac{R}{R_{V}}\right)  ^{3}.
\end{equation}
where the lower (plus) sign is for the under-saturated solution and the upper
(minus) sign for the super-saturated solution. In the case of the
undersaturated solution, the stationary distribution is then%
\begin{equation}
P\left(  R\right)  =\mathcal{N}R^{3/2}\exp\left(  -\beta\Delta F\left(
R\right)  \right)  \left(  1+O\left(  \frac{R}{R_{V}}\right)  \right)
\end{equation}
where $\mathcal{N}$ is a normalization factor. This is the CNT expression for
the distribution of clusters in the undersaturated solution.

Finally, the (weak-noise limit of the ) Fokker-Planck equation is%
\begin{equation}
\frac{\partial}{\partial t}P_{t}\left(  R\right)  =D\frac{\partial}{\partial
R}\left(  g_{RR}^{-1}\left(  R\right)  \frac{\partial\beta F\left(  R\right)
}{\partial R}+g_{RR}^{-1}\left(  R\right)  \frac{\partial}{\partial R}\right)
P_{t}\left(  R\right)  .
\end{equation}
This describes $P_{t}\left(  R\right)  $, the probability for a given cluster
to have radius $R$, but it can easily be related to $C(N)$, the concentration
of clusters containing $N$ moleculeso\cite{lutsko2013b}, which then satisfies
a similar equation (again, keeping only terms appropriate to the weak-noise
limit),
\begin{equation}
\frac{\partial}{\partial t}C_{t}\left(  N\right)  =\frac{\partial}{\partial
N}\left(  Dg_{NN}^{-1}\left(  N\right)  \frac{\partial\beta F\left(  N\right)
}{\partial N}+Dg_{NN}^{-1}\left(  N\right)  \frac{\partial}{\partial
N}\right)  C_{t}\left(  N\right)  . \label{z}%
\end{equation}
with%
\begin{equation}
g_{NN}\left(  N\right)  =g_{RR}\left(  R\right)  \left(  \frac{dR}{dN}\right)
^{2}=\frac{\left(  \rho_{0}-\overline{\rho}\right)  ^{2}}{4\pi R\rho_{0}%
^{2}\overline{\rho}}.
\end{equation}
For a weak solution, $\rho_{0}\gg\overline{\rho}$, one then has that%
\begin{equation}
Dg_{NN}^{-1}\left(  N\right)  \simeq4\pi RD\overline{\rho}.
\end{equation}
In CNT, the same result is derived for the case of diffusion-limited
homogeneous nucleation in which case Eq.(\ref{z}) is recognized as the
Zeldovich equation with the "attachment rate" $f_{N}=Dg_{NN}^{-1}\left(
N\right)  $ for which the expression given here agrees with that derived in
CNT (see e.g. Ref.\cite{Kashchiev}, Eq. 10.18) except that the latter includes
a heuristic coefficient (the "sticking probability") inserted by hand. Thus,
the theory recovers the well-known results of CNT in this limit.

\subsection{Generalization: CNT with moving clusters}

The CNT model can be generalized by allowing the clusters to move. In this
case, we begin with%
\begin{equation}
\rho\left(  \mathbf{r};\mathbf{x}\right)  =\rho_{0}\Theta\left(  R-\left\vert
\mathbf{r-\Delta}\right\vert \right)  +\rho_{\infty}\left(  \mathbf{x}\right)
\Theta\left(  \left\vert \mathbf{r-\Delta}\right\vert -R\right)
\end{equation}
so that the parameters are the radius, $R,$ and the location of the center of
the cluster, $\mathbf{\Delta}$ . Solving the Poisson equation and neglecting
finite size terms (e.g. of order $R/R_{T}$ and $\mathbf{\Delta}/R_{T}$ ), one
finds%
\begin{align}
g_{RR}  &  =4\pi\frac{\left(  \rho_{0}-\rho_{\infty}\right)  ^{2}}%
{\rho_{\infty}}R^{3}\\
g_{\Delta_{i}\Delta_{i}}  &  =\frac{4\pi}{3}\frac{\left(  \rho_{0}%
-\rho_{\infty}\right)  ^{2}}{\rho_{0}+2\rho_{\infty}}R^{3}\nonumber
\end{align}
and all off diagonal terms vanish. Since the free energy is independent of
$\Delta$ , the center of mass just undergoes brownian motion with diffusion
constant $D\frac{3}{4\pi R^{3}}\frac{\rho_{0}+2\rho_{\infty}}{\left(  \rho
_{0}-\rho_{\infty}\right)  ^{2}}$.

\subsection{Generalization:\ Ellipsoidal Clusters}

The surface of an ellipsoid with axes aligned along the Cartesian directions
is specified by three parameters, $a_{1},a_{2}$ and $a_{3}$ as%
\begin{equation}
\frac{x^{2}}{a_{1}^{2}}+\frac{y^{2}}{a_{2}^{2}}+\frac{z^{2}}{a_{3}^{2}}=1.
\end{equation}
It includes the oblate spheroid ($a_{1}=a_{2}>a_{3}$), the prolate spheroid
($a_{1}>a_{2}=a_{3}$) and the sphere ($a_{1}=a_{2}=a_{3}$) as obvious special
cases. Somewhat more convenient parameters are the average radius and the
eccentricities defined (assuming $a_{1}>a_{2}$, $a_{3}$) respectively as%
\begin{align}
R  &  =\left(  a_{1}a_{2}a_{3}\right)  ^{1/3}\\
x_{1}  &  =\varepsilon_{1}^{2}=\frac{a_{1}^{2}-a_{2}^{2}}{a_{1}^{2}%
}\nonumber\\
x_{2}  &  =\varepsilon_{2}^{2}=\frac{a_{1}^{2}-a_{3}^{2}}{a_{1}^{2}}.\nonumber
\end{align}
so that the radius is the overall measure of size of the ellipsoid and the
eccentricities are measures of the shape. Indeed, the volume and surface area
are%
\begin{align}
V\left(  R\right)   &  =\frac{4\pi}{3}R^{3}\\
S\left(  R,x_{1},x_{2}\right)   &  =2\pi R^{2}\frac{1}{\left(  1-\varepsilon
_{1}^{2}\right)  ^{1/3}\left(  1-\varepsilon_{2}^{2}\right)  ^{1/3}}\left\{
\begin{array}
[c]{c}%
1-\varepsilon_{2}^{2}+\sqrt{1-\varepsilon_{1}^{2}}\varepsilon_{2}E\left(
\arcsin\varepsilon_{2};\sqrt{\frac{\varepsilon_{2}^{2}-\varepsilon_{1}^{2}%
}{\varepsilon_{2}^{2}\left(  1-\varepsilon_{1}^{2}\right)  }}\right) \\
+\frac{\sqrt{1-\varepsilon_{1}^{2}}\left(  1-\varepsilon_{2}^{2}\right)
}{\varepsilon_{2}}F\left(  \arcsin\varepsilon_{2};\sqrt{\frac{\varepsilon
_{2}^{2}-\varepsilon_{1}^{2}}{\varepsilon_{2}^{2}\left(  1-\varepsilon_{1}%
^{2}\right)  }}\right)
\end{array}
\right\} \nonumber
\end{align}
where $E\left(  \phi,k\right)  $ and $F\left(  \phi,k\right)  $ are incomplete
elliptic integrals of the first and second kind. For small eccentricities, the
area can be expanded to get%
\begin{equation}
S\left(  R,\varepsilon_{1},\varepsilon_{2}\right)  =4\pi R^{2}\left(
1+\frac{2}{45}\left(  \varepsilon_{1}^{4}+\varepsilon_{2}^{4}-\varepsilon
_{1}^{2}\varepsilon_{2}^{2}\right)  +...\right)  .
\end{equation}
The capillary model for the cluster is
\begin{equation}
\rho\left(  \mathbf{r};a_{1},a_{2},a_{3}\right)  =\rho_{l}\Theta\left(
1-\left(  \frac{x^{2}}{a_{1}^{2}}+\frac{y^{2}}{a_{2}^{2}}+\frac{z^{2}}%
{a_{3}^{2}}\right)  \right)  +\rho_{v}\Theta\left(  \frac{x^{2}}{a_{1}^{2}%
}+\frac{y^{2}}{a_{2}^{2}}+\frac{z^{2}}{a_{3}^{2}}-1\right)
\end{equation}
and for the free energy,
\begin{equation}
\Delta F\left(  R,\varepsilon_{1},\varepsilon_{2}\right)  =V\left(  R\right)
\left(  f_{l}-f_{v}\right)  +S\left(  R,\varepsilon_{1},\varepsilon
_{2}\right)  \gamma.
\end{equation}
The only other element needed is the metric. This is obtained by a
straightforward but somewhat involved calculation. The details and exact
results are given in the SI. Here, I\ will only give the result for two
limits.\ First, in the weak solution appoximation ($\rho_{v}/\rho_{l}\ll1$),
the leading order contributions to the metric are%
\begin{align}
g_{RR}  &  \mathbf{=}\frac{4\pi\rho_{l}^{2}}{\rho_{v}}\frac{\left(
1-\varepsilon_{1}^{2}\right)  ^{1/6}\left(  1-\varepsilon_{2}^{2}\right)
^{1/6}}{\varepsilon_{1}}F\left(  \arcsin\left(  \varepsilon_{2}\right)
|\frac{\varepsilon_{1}^{2}}{\varepsilon_{2}^{2}}\right)  R^{3}\\
g_{\varepsilon_{1}\varepsilon_{1}}  &  \mathbf{=}\frac{4\pi\rho_{l}}{135}%
\frac{\varepsilon_{1}^{2}}{\left(  1-\varepsilon_{1}^{2}\right)  ^{7/3}\left(
1-\varepsilon_{2}^{2}\right)  ^{1/3}}\allowbreak\left(  6-4\varepsilon_{1}%
^{2}-\varepsilon_{2}^{2}\right)  R^{5}\nonumber\\
g_{\varepsilon_{1}\varepsilon_{2}}  &  =g_{\varepsilon_{2}\varepsilon_{1}%
}\mathbf{=}\frac{4\pi\rho_{l}}{135}\frac{\varepsilon_{1}\varepsilon_{2}%
}{\left(  1-\varepsilon_{1}^{2}\right)  ^{4/3}\left(  1-\varepsilon_{2}%
^{2}\right)  ^{4/3}}\allowbreak\allowbreak\left(  3-2\varepsilon_{1}%
^{2}-2\varepsilon_{2}^{2}\right)  R^{5}\nonumber\\
g_{\varepsilon_{2}\varepsilon_{2}}  &  \mathbf{=}\frac{4\pi\rho_{l}}{135}%
\frac{\varepsilon_{2}^{2}}{\left(  1-\varepsilon_{1}^{2}\right)  ^{1/3}\left(
1-\varepsilon_{2}^{2}\right)  ^{7/3}}\allowbreak\allowbreak\left(
6-\varepsilon_{1}^{2}-4\varepsilon_{2}^{2}\right)  R^{5}\nonumber
\end{align}
and all other elements are zero. The determinant is
\begin{equation}
\det g=-\frac{64\pi^{3}}{2025}\frac{\rho_{l}^{4}}{\rho_{v}}R^{13}%
\frac{\varepsilon_{1}^{2}\varepsilon_{2}}{\left(  1-\varepsilon_{1}%
^{2}\right)  ^{\frac{5}{2}}\left(  1-\varepsilon_{2}^{2}\right)  ^{\frac{5}%
{2}}}\allowbreak\left(  3-2\varepsilon_{1}^{2}-2\varepsilon_{2}^{2}%
+\varepsilon_{1}^{2}\varepsilon_{2}^{2}\right)  F\left(  \arcsin\left(
\varepsilon_{2}\right)  |\frac{\varepsilon_{1}^{2}}{\varepsilon_{2}^{2}%
}\right)
\end{equation}

To lowest order in the eccentricities, the equilibrium distribution will be
\begin{equation}
P\sim R^{13/2}\varepsilon_{1}\varepsilon_{2}\exp\left(  -V\left(  R\right)
\left(  \beta f_{l}-\beta f_{v}\right)  -S\left(  R\right)  \beta\gamma
-\frac{2}{45}\beta\gamma S\left(  R\right)  \left(  \varepsilon_{1}%
^{4}+\varepsilon_{2}^{4}-\varepsilon_{1}^{2}\varepsilon_{2}^{2}\right)
\right)
\end{equation}
The critical cluster occurs at
\begin{equation}
\varepsilon_{1}=\varepsilon_{2}=0
\end{equation}
and with the critical radius and energy barrier calculated for a spherical
cluster. The excess free energy can be written as
\begin{equation}
\beta\Delta F=\beta\Delta F_{c}\left[  -2\left(  \frac{R}{R_{c}}\right)
^{3}+3\left(  \frac{R}{R_{c}}\right)  ^{2}+\frac{2}{15}\left(  \frac{R}{R_{c}%
}\right)  ^{2}\left(  \varepsilon_{1}^{4}+\varepsilon_{2}^{4}-\varepsilon
_{1}^{2}\varepsilon_{2}^{2}\right)  \right]
\end{equation}
Thus fluctuations in the eccecentricites are strongly damped suggesting an
alternative expansion wherein the eccentricities are treated as small
parameters and the densities are unconstrained. The lowest order results in
this case are%
\begin{align}
\overline{\overline{g}}_{RR}  &  =\frac{4\pi\left(  \rho_{0}-\rho_{1}\right)
^{2}}{\rho_{1}}R^{3}\left(  1+O\left(  \varepsilon_{1}^{4},\varepsilon_{2}%
^{4},\varepsilon_{1}^{2}\varepsilon_{2}^{2}\right)  \right) \\
\overline{\overline{g}}_{\varepsilon_{1}\varepsilon_{1}}  &  =\left(
2\varepsilon_{1}\right)  ^{2}\frac{8\pi}{45}\frac{\left(  \rho_{0}-\rho
_{1}\right)  ^{2}}{2\rho_{0}+3\rho_{1}}R^{5}\left(  1+\frac{1}{21}\frac
{1}{2\rho_{0}+3\rho_{1}}\left(  10\left(  7\rho_{0}+12\rho_{1}\right)
\varepsilon_{1}^{2}+\left(  7\rho_{0}+3\rho_{1}\right)  \varepsilon_{2}%
^{2}\right)  \right) \nonumber\\
\overline{\overline{g}}_{\varepsilon_{1}\varepsilon_{2}}  &  =-\left(
2\varepsilon_{1}\right)  \left(  2\varepsilon_{2}\right)  \frac{4\pi}{45}%
\frac{\left(  \rho_{0}-\rho_{1}\right)  ^{2}}{2\rho_{0}+3\rho_{1}}R^{5}\left(
1+\frac{1}{21}\frac{28\rho_{0}+57\rho_{1}}{2\rho_{0}+3\rho_{1}}\left(
\varepsilon_{1}^{2}+\varepsilon_{2}^{2}\right)  \right) \nonumber\\
\overline{\overline{g}}_{\varepsilon_{2}\varepsilon_{2}}  &  =\left(
2\varepsilon_{2}\right)  ^{2}\frac{8\pi}{45}\frac{\left(  \rho_{0}-\rho
_{1}\right)  ^{2}}{2\rho_{0}+3\rho_{1}}R^{5}\left(  1+\frac{1}{21}\frac
{1}{2\rho_{0}+3\rho_{1}}\left(  \left(  7\rho_{0}+3\rho_{1}\right)
\varepsilon_{1}^{2}+10\left(  7\rho_{0}+12\rho_{1}\right)  \varepsilon_{2}%
^{2}\right)  \right) \nonumber
\end{align}
and,
\begin{equation}
\det\overline{\overline{\mathbf{g}}}=\left(  2\varepsilon_{1}\right)
^{2}\left(  2\varepsilon_{2}\right)  ^{2}\frac{\left(  4\pi\left(  \rho
_{0}-\rho_{1}\right)  ^{2}\right)  ^{3}}{\rho_{1}\left(  2\rho_{0}+3\rho
_{1}\right)  ^{2}}R^{13}\frac{1}{675}\left[  1+2\varepsilon_{1}^{2}%
+2\varepsilon_{2}^{2}\right]
\end{equation}
The marginal distribution for the radius is
\begin{equation}
P\sim R^{9/2}\exp\left(  -\beta\Delta F_{c}\left[  -2\left(  \frac{R}{R_{c}%
}\right)  ^{3}+3\left(  \frac{R}{R_{c}}\right)  ^{2}\right]  \right)
\end{equation}
which differs from that one would get by freezing out the eccentricities,
\begin{equation}
P\sim R^{3/2}\exp\left(  -\beta\Delta F_{c}\left[  -2\left(  \frac{R}{R_{c}%
}\right)  ^{3}+3\left(  \frac{R}{R_{c}}\right)  ^{2}\right]  \right)
\end{equation}
and the determinant of the Hessian is
\begin{equation}
\det\beta\Delta F_{c}\left(
\begin{array}
[c]{ccc}%
-6 & 0 & 0\\
0 & \frac{4}{15} & -\frac{2}{15}\\
0 & -\frac{2}{15} & \frac{4}{15}%
\end{array}
\right)  =-\frac{8}{25}\left(  \beta\Delta F_{c}\right)  ^{3}%
\end{equation}
The unstable eigenvalue of $gH$ is
\begin{equation}
\frac{24\pi\left(  \rho_{0}-\rho_{1}\right)  ^{2}}{\rho_{1}}R_{c}^{5}%
\beta\Delta F_{c}%
\end{equation}

\section{Conclusions}

Classical nucleation theory has long provided not only a mathematical model
for nucleation allowing one to estimate nucleation rates and other physically
interesting quantities, but also the language used to discuss nucleation:
concepts such as the critical nucleus, the competition between surface tension
and bulk free energy differences, etc. However, it is also severly limited in
applicability due to the underlying assumptions of spherical clusters that are
large compared to the growth units, slow growth and many others. As attention
is increasingly drawn to problems that violate those assumptions - i.e.
nanoscale processes, multistep nucleation - the obvious alternative is to turn
to a more fundamental description such as kinetic theory or fluctuating
hydrodynamics but the price paid is to lose contact with the familiar
phenomenology and language. The goal of this paper has been to develop a
bridge between these two levels of description that allows for the development
of post-CNT models that are nevertheless grounded in the more microscopic approaches.

The structure of such a model, stochastic models with a deterministic driving
force based on free energy gradients and a fluctuating force with amplitude
determined by a fluctuation-dissipation relation - could be guessed and is
enough for the weak-noise limit. The requirement of covariance supplies
additional information needed to generalize beyond weak noise. What is
missing, and what cannot be guessed, are the kinitic coefficients - what has
been called here, the metric - which govern the kinetics of the process. In
the domain of CNT, the kinetics are usually thought to be of secondary
importance since the exponential dependence of the nucleation rate on the free
energy barrier dominates practical calculations. However, outside this domain
the barriers become smaller and kinetics becomes more important. The main
contribution of this work has been to clarify how the metric should be
determined based on the parameterization of the density and how to then
construct a self-consistent model.

The general framework has been illustrated first by recovering previous
results for spherically-symmetric systems, which include CNT when reduced to a
single order parameter. Then, two novel generalizations were developed - one
for moving clusters and the second for ellipsoidal clusters. The latter in
particular could be used as a basis for incorporating the effects of shear on
nucleation. In any case, the goal here was not to delve too much into
applications but, rather, to illustrate how such models can be developed with
minimal heuristic input.

\begin{acknowledgments}
This work was supported in part by the European Space Agency under contract
number ESA AO-2004-070.
\end{acknowledgments}

%

\appendix

\section{Appendix:\ Proof of Eq.(\ref{weak})\label{AppComplete}}

Using the completeness relation,
\begin{align}
\frac{\partial\rho\left(  \mathbf{r};\mathbf{x}\right)  }{\partial x^{\alpha}%
}\frac{\delta x^{\alpha}}{\delta\rho\left(  \mathbf{r}^{\prime};\mathbf{x}%
\right)  }  &  =\delta\left(  \mathbf{r-r}^{\prime}\right) \\
\int\frac{\delta x^{\alpha}}{\delta\rho\left(  \mathbf{r};\mathbf{x}\right)
}\frac{\partial\rho\left(  \mathbf{r};\mathbf{x}\right)  }{\partial x^{\beta}%
}d\mathbf{r}  &  =\delta_{\beta}^{\alpha}\nonumber
\end{align}
the last term on the right in Eq.(\ref{l1}) could be reformulated as
\begin{align}
&  \int\frac{\partial\beta F\left(  \mathbf{x}\right)  }{\partial x^{\alpha}%
}\frac{\delta x^{\alpha}}{\delta\rho\left(  \mathbf{r};\mathbf{x}\right)
}\left(  -\mathbf{\nabla}\cdot\rho\left(  \mathbf{r};\mathbf{x}\right)
\mathbf{\nabla}\right)  \frac{\delta x^{\beta}}{\delta\rho\left(
\mathbf{r};\mathbf{x}\right)  }\frac{\partial\beta F\left(  \mathbf{x}\right)
}{\partial x^{\beta}}d\mathbf{r}\\
&  \mathbf{=}\frac{\partial\beta F\left(  \mathbf{x}\right)  }{\partial
x^{\alpha}}\left(  \int\frac{\partial\rho\left(  \mathbf{r};\mathbf{x}\right)
}{\partial x^{\beta}}\left(  -\mathbf{\nabla}\cdot\rho\left(  \mathbf{r}%
;\mathbf{x}\right)  \mathbf{\nabla}\right)  ^{-1}\frac{\partial\rho\left(
\mathbf{r};\mathbf{x}\right)  }{\partial x^{\alpha}}d\mathbf{r}\right)
^{-1}\frac{\partial\beta F\left(  \mathbf{x}_{t}\right)  }{\partial
x_{t}^{\beta}}\nonumber\\
&  \mathbf{=}\frac{\partial\beta F\left(  \mathbf{x}\right)  }{\partial
x^{\alpha}}g^{\alpha\beta}\left(  \mathbf{x}\right)  \frac{\partial\beta
F\left(  \mathbf{x}\right)  }{\partial x^{\beta}}\nonumber
\end{align}
The equation between the first and second lines also requires the completeness
relations since if
\begin{equation}
K^{\alpha\beta}\left(  \mathbf{x}\right)  =\int\frac{\delta x^{\alpha}}%
{\delta\rho\left(  \mathbf{r}^{\prime};\mathbf{x}\right)  }\left(
-\mathbf{\nabla}^{\prime}\cdot\rho\left(  \mathbf{r}^{\prime};\mathbf{x}%
\right)  \mathbf{\nabla}^{\prime}\right)  \frac{\delta x^{\beta}}{\delta
\rho\left(  \mathbf{r}^{\prime};\mathbf{x}\right)  }d\mathbf{r}^{\prime}%
\end{equation}
then%
\begin{align}
\frac{\partial\rho\left(  \mathbf{r};\mathbf{x}\right)  }{\partial x^{\alpha}%
}K^{\alpha\beta}\left(  \mathbf{x}\right)   &  =\int\frac{\partial\rho\left(
\mathbf{r};\mathbf{x}\right)  }{\partial x^{\alpha}}\frac{\delta x^{\alpha}%
}{\delta\rho\left(  \mathbf{r}^{\prime};\mathbf{x}\right)  }\left(
-\mathbf{\nabla}^{\prime}\cdot\rho\left(  \mathbf{r}^{\prime};\mathbf{x}%
\right)  \mathbf{\nabla}^{\prime}\right)  \frac{\delta x^{\beta}}{\delta
\rho\left(  \mathbf{r}^{\prime};\mathbf{x}\right)  }d\mathbf{r}^{\prime}\\
&  =\int\delta\left(  \mathbf{r-r}^{\prime}\right)  \left(  -\mathbf{\nabla
}^{\prime}\cdot\rho\left(  \mathbf{r}^{\prime};\mathbf{x}\right)
\mathbf{\nabla}^{\prime}\right)  \frac{\delta x^{\beta}}{\delta\rho\left(
\mathbf{r}^{\prime};\mathbf{x}\right)  }d\mathbf{r}^{\prime}\nonumber\\
&  =\left(  -\mathbf{\nabla}\cdot\rho\left(  \mathbf{r};\mathbf{x}\right)
\mathbf{\nabla}\right)  \frac{\delta x^{\beta}}{\delta\rho\left(
\mathbf{r};\mathbf{x}\right)  }\nonumber
\end{align}
and%
\begin{align}
\delta_{\gamma}^{\beta}  &  =\int\frac{\partial\rho\left(  \mathbf{r}%
;\mathbf{x}\right)  }{\partial x^{\gamma}}\frac{\delta x_{t}^{\beta}}%
{\delta\rho\left(  \mathbf{r};\mathbf{x}\right)  }d\mathbf{r}\\
&  =\left(  \int\frac{\partial\rho\left(  \mathbf{r};\mathbf{x}\right)
}{\partial x^{\gamma}}\left(  -\mathbf{\nabla}\cdot\rho\left(  \mathbf{r}%
;\mathbf{x}\right)  \mathbf{\nabla}\right)  ^{-1}\frac{\partial\rho\left(
\mathbf{r};\mathbf{x}\right)  }{\partial x^{\alpha}}d\mathbf{r}\right)
K^{\alpha\beta}\nonumber\\
&  =g_{\gamma\alpha}\left(  \mathbf{x}\right)  K^{\alpha\beta}\left(
\mathbf{x}\right)  .\nonumber
\end{align}
So, $K^{\alpha\beta}=g^{\alpha\beta}$ as claimed. Inserting this result into
Eq.(\ref{l1}) gives for the weak-noise Lagrangian of the paramaterized paths%
\begin{align}
\mathcal{L}\left(  \mathbf{x}_{t}\right)   &  =\frac{dx_{t}^{\alpha}}%
{dt}g_{\alpha\beta}\left(  \mathbf{x}_{t}\right)  \frac{dx_{t}^{\beta}}%
{dt}+2\frac{dx_{t}^{\alpha}}{dt}\frac{\partial\beta F\left(  \mathbf{x}%
_{t}\right)  }{\partial x_{t}^{\alpha}}+\frac{\partial\beta F\left(
\mathbf{x}_{t}\right)  }{\partial x_{t}^{\alpha}}g^{\alpha\beta}\frac
{\partial\beta F\left(  \mathbf{x}_{t}\right)  }{\partial x_{t}^{\beta}}\\
&  =\left(  \frac{dx_{t}^{\alpha}}{dt}+g^{\alpha\beta}\left(  \mathbf{x}%
_{t}\right)  \frac{\partial\beta F\left(  \mathbf{x}_{t}\right)  }{\partial
x_{t}^{\beta}}\right)  g_{\alpha\gamma}\left(  \mathbf{x}_{t}\right)  \left(
\frac{dx_{t}^{\gamma}}{dt}+g^{\gamma\sigma}\left(  \mathbf{x}_{t}\right)
\frac{\partial\beta F\left(  \mathbf{x}_{t}\right)  }{\partial x_{t}^{\sigma}%
}\right) \nonumber
\end{align}

\section{Deriving the SDE\label{AppDeriv}}

The SDE for the density is
\begin{equation}
\frac{\partial\rho_{t}\left(  \mathbf{r}\right)  }{\partial t}=D\mathbf{\nabla
}\cdot\rho_{t}\left(  \mathbf{r}\right)  \mathbf{\nabla}\frac{\delta\beta
F\left[  \rho_{t}\right]  }{\delta\rho_{t}\left(  \mathbf{r}\right)
}+\mathbf{\nabla}\cdot\sqrt{2D\rho_{t}\left(  \mathbf{r}\right)  }\mathbf{\xi
}_{t}\left(  \mathbf{r}\right)
\end{equation}
and we use the Stratanovich interpretation (recall that this model is actually
Ito-Statonovich equivalent). If the density can be represented as%
\begin{equation}
\rho_{t}\left(  \mathbf{r}\right)  =\rho\left(  \mathbf{r;x}_{t}\right)
\end{equation}
then%
\begin{equation}
\frac{\partial\rho\left(  \mathbf{r;x}_{t}\right)  }{\partial x_{t}^{\alpha}%
}\frac{dx_{t}^{\alpha}}{dt}=D\mathbf{\nabla}\cdot\rho\left(  \mathbf{r;x}%
_{t}\right)  \mathbf{\nabla}\left(  \frac{\delta\beta F\left[  \rho\right]
}{\delta\rho\left(  \mathbf{r}\right)  }\right)  _{\rho\left(  \mathbf{r}%
\right)  =\rho\left(  \mathbf{r;x}_{t}\right)  }+\mathbf{\nabla}\cdot
\sqrt{2D\rho\left(  \mathbf{r;x}_{t}\right)  }\mathbf{\xi}_{t}\left(
\mathbf{r}\right)  .
\end{equation}
Multiplying through by the appropriate operator gives%
\begin{align}
\frac{\partial\rho\left(  \mathbf{r;x}_{t}\right)  }{\partial x_{t}^{\beta}%
}\left(  \mathbf{\nabla}\cdot\rho\left(  \mathbf{r;x}_{t}\right)
\mathbf{\nabla}\right)  ^{-1}\frac{\partial\rho\left(  \mathbf{r;x}%
_{t}\right)  }{\partial x_{t}^{\alpha}}\frac{dx_{t}^{\alpha}}{dt}  &
=D\frac{\partial\rho\left(  \mathbf{r;x}_{t}\right)  }{\partial x_{t}^{\beta}%
}\frac{\delta\beta F\left[  \rho_{t}\right]  }{\delta\rho_{t}\left(
\mathbf{r}\right)  }\\
&  +\frac{\partial\rho\left(  \mathbf{r;x}_{t}\right)  }{\partial x_{t}%
^{\beta}}\left(  \mathbf{\nabla}\cdot\rho\left(  \mathbf{r;x}_{t}\right)
\mathbf{\nabla}\right)  ^{-1}\mathbf{\nabla}\cdot\sqrt{2D\rho\left(
\mathbf{r;x}_{t}\right)  }\mathbf{\xi}_{t}\left(  \mathbf{r}\right) \nonumber
\end{align}
Integrating over the spatial coordinates yields%
\begin{equation}
-g_{\beta\alpha}\left(  \mathbf{x}_{t}\right)  \frac{dx_{t}^{\alpha}}%
{dt}=D\frac{\partial\beta F\left(  \mathbf{x}_{t}\right)  }{\partial
x_{t}^{\beta}}+g_{\alpha\beta}\left(  \mathbf{x}_{t}\right)  \int_{V}q^{\beta
}\left(  \mathbf{r}\right)  \mathbf{\xi}_{t}\left(  \mathbf{r}\right)
d\mathbf{r}%
\end{equation}
where we have identified the metric in the space of parameters,
\begin{equation}
g_{\beta\alpha}\left(  \mathbf{x}_{t}\right)  =-\int_{V}\frac{\partial
\rho\left(  \mathbf{r;x}_{t}\right)  }{\partial\mathbf{x}_{t}^{\beta}}\left(
\mathbf{\nabla}\cdot\rho\left(  \mathbf{r;x}_{t}\right)  \mathbf{\nabla
}\right)  ^{-1}\frac{\partial\rho\left(  \mathbf{r;x}_{t}\right)  }%
{\partial\mathbf{x}_{t}^{\alpha}}d\mathbf{r,}%
\end{equation}
and the parameterized free energy%
\begin{equation}
F\left(  \mathbf{x}\right)  =\left.  \beta F\left[  \rho\right]  \right\vert
_{\rho\left(  \mathbf{r}\right)  =\rho\left(  \mathbf{r;x}\right)  },
\end{equation}
and the noise amplitude is%
\begin{equation}
q_{a}^{\beta}\left(  \mathbf{r}\right)  =g^{\beta\gamma}\left(  \mathbf{x}%
_{t}\right)  \frac{\partial\rho\left(  \mathbf{r;}x_{t}\right)  }{\partial
x_{t}^{\gamma}}\left(  \mathbf{\nabla}\cdot\rho\left(  \mathbf{r;x}%
_{t}\right)  \mathbf{\nabla}\right)  ^{-1}\frac{\partial}{\partial r_{a}}%
\sqrt{2D\rho\left(  \mathbf{r;x}_{t}\right)  }.
\end{equation}
Writing this as
\begin{equation}
\frac{dx_{t}^{\alpha}}{dt}=-Dg^{\alpha\beta}\left(  \mathbf{x}_{t}\right)
\frac{\partial\beta F\left(  \mathbf{x}_{t}\right)  }{\partial x_{t}^{\beta}%
}+\int_{V}q^{\beta}\left(  \mathbf{r}\right)  \mathbf{\xi}_{t}\left(
\mathbf{r}\right)  d\mathbf{r}%
\end{equation}
we note that the equivalent Fokker-Planck equation is
\begin{equation}
\frac{\partial}{\partial t}P_{t}\left(  \mathbf{x}\right)  =\frac{\partial
}{\partial x^{\alpha}}\left(  g^{\alpha\beta}\left(  \mathbf{x}\right)
D\frac{\partial\beta F\left(  \mathbf{x}\right)  }{\partial x^{\beta}}%
+\frac{1}{2}q_{a}^{\alpha}\left(  \mathbf{r}\right)  \frac{\partial}{\partial
x^{\beta}}q_{a}^{\beta}\left(  \mathbf{r}^{\prime}\right)  \delta\left(
\mathbf{r}-\mathbf{r}^{\prime}\right)  \right)  P_{t}\left(  \mathbf{x}%
\right)
\end{equation}
(with an implied integration over the spatial coordinate). This can be written
in anti-Ito form as
\begin{equation}
\frac{\partial}{\partial t}P_{t}\left(  \mathbf{x}\right)  =\frac{\partial
}{\partial x^{\alpha}}\left(  g^{\alpha\beta}\left(  \mathbf{x}\right)
D\frac{\partial\beta F\left(  \mathbf{x}\right)  }{\partial x^{\beta}}%
+\frac{1}{2}q_{a}^{\alpha}\left(  \mathbf{r}\right)  \left(  \frac{\partial
}{\partial x^{\beta}}q_{a}^{\beta}\left(  \mathbf{r}^{\prime}\right)
\delta\left(  \mathbf{r}-\mathbf{r}^{\prime}\right)  \right)  +\frac{1}%
{2}q_{a}^{\alpha}\left(  \mathbf{r}\right)  q_{a}^{\beta}\left(
\mathbf{r}^{\prime}\right)  \delta\left(  \mathbf{r}-\mathbf{r}^{\prime
}\right)  \frac{\partial}{\partial x^{\beta}}\right)  P_{t}\left(
\mathbf{x}\right)
\end{equation}
Using the results below,%
\begin{equation}
\frac{\partial}{\partial t}P_{t}\left(  \mathbf{x}\right)  =\frac{\partial
}{\partial x^{\alpha}}\left(  g^{\alpha\beta}\left(  \mathbf{x}\right)
D\frac{\partial\beta F\left(  \mathbf{x}\right)  }{\partial x^{\beta}}%
-\frac{1}{2}Dg^{\alpha\gamma}\left(  \mathbf{x}\right)  \frac{1}%
{\det\mathbf{g}\left(  \mathbf{x}\right)  }\frac{\partial}{\partial x^{\gamma
}}\det\mathbf{g}\left(  \mathbf{x}\right)  +\frac{1}{2}DA^{\alpha}\left(
\mathbf{x}\right)  +Dg^{\alpha\beta}\left(  \mathbf{x}\right)  \frac{\partial
}{\partial x^{\beta}}\right)  P_{t}\left(  \mathbf{x}\right)
\end{equation}
with%
\begin{align}
A^{\alpha}\left(  \mathbf{x}\right)   &  =\left(  g^{\alpha\beta}\left(
\mathbf{x}\right)  g^{\gamma\sigma}\left(  \mathbf{x}\right)  -g^{\alpha
\gamma}\left(  \mathbf{x}\right)  g^{\beta\sigma}\left(  \mathbf{x}\right)
\right) \\
&  \times\int_{V}d\mathbf{r\;}\frac{\partial\rho\left(  \mathbf{r;x}\right)
}{\partial x^{\beta}}\left(  \mathbf{\nabla}\left[  \left(  \mathbf{\nabla
}\cdot\rho\left(  \mathbf{r;x}\right)  \mathbf{\nabla}\right)  ^{-1}%
\frac{\partial\rho\left(  \mathbf{r;}x\right)  }{\partial x^{\gamma}}\right]
\right)  \cdot\left(  \mathbf{\nabla}\left[  \left(  \mathbf{\nabla}\cdot
\rho\left(  \mathbf{r;x}\right)  \mathbf{\nabla}\right)  ^{-1}\frac
{\partial\rho\left(  \mathbf{r;}x\right)  }{\partial x^{\sigma}}\right]
\right) \nonumber
\end{align}
Note that in this case, the former can be written as
\begin{align}
\frac{\partial}{\partial t}P_{t}\left(  \mathbf{x}\right)   &  =D\frac
{\partial}{\partial x^{\alpha}}\left(  g^{\alpha\beta}\left(  \mathbf{x}%
\right)  \frac{\partial\beta F\left(  \mathbf{x}\right)  }{\partial x^{\beta}%
}+g^{\alpha\beta}\left(  \mathbf{x}\right)  \sqrt{\det\mathbf{g}\left(
\mathbf{x}\right)  }\frac{\partial}{\partial x^{\beta}}\frac{1}{\sqrt
{\det\mathbf{g}\left(  \mathbf{x}\right)  }}+\frac{1}{2}A^{\alpha}\left(
\mathbf{x}\right)  +g^{\alpha\beta}\left(  \mathbf{x}\right)  \frac{\partial
}{\partial x^{\beta}}\right)  P_{t}\left(  \mathbf{x}\right) \\
&  =D\frac{\partial}{\partial x^{\alpha}}\left(  g^{\alpha\beta}\left(
\mathbf{x}\right)  \frac{\partial\beta F\left(  \mathbf{x}\right)  }{\partial
x^{\beta}}+\frac{1}{2}A^{\alpha}\left(  \mathbf{x}\right)  +g^{\alpha\beta
}\left(  \mathbf{x}\right)  \sqrt{\det\mathbf{g}\left(  \mathbf{x}\right)
}\frac{\partial}{\partial x^{\beta}}\frac{1}{\sqrt{\det\mathbf{g}\left(
\mathbf{x}\right)  }}\right)  P_{t}\left(  \mathbf{x}\right) \\
&  =D\frac{\partial}{\partial x^{\alpha}}\left(  \frac{1}{2}A^{\alpha}\left(
\mathbf{x}\right)  +g^{\alpha\beta}\left(  \mathbf{x}\right)  \sqrt
{\det\mathbf{g}\left(  \mathbf{x}\right)  }\exp\left(  -\beta F\left(
\mathbf{x}\right)  \right)  \frac{\partial}{\partial x^{\beta}}\frac{1}%
{\sqrt{\det\mathbf{g}\left(  \mathbf{x}\right)  }}\exp\left(  \beta F\left(
\mathbf{x}\right)  \right)  \right)  P_{t}\left(  \mathbf{x}\right)
\end{align}
showing that if $A^{\alpha}$ is neglected, then a stationary solution is
\begin{equation}
P\left(  \mathbf{x}\right)  =\sqrt{\det\mathbf{g}\left(  \mathbf{x}\right)
}\exp\left(  -\beta F\left(  \mathbf{x}\right)  \right)  .
\end{equation}
The Fokker-Planck equation is equivalent to the anti-Ito sde%
\begin{equation}
\frac{dx_{t}^{\alpha}}{dt}=-Dg^{\alpha\beta}\left(  \mathbf{x}_{t}\right)
\frac{\partial\beta F\left(  \mathbf{x}_{t}\right)  }{\partial x_{t}^{\beta}%
}+\frac{1}{2}Dg^{\alpha\beta}\left(  \mathbf{x}_{t}\right)  \frac{1}%
{\det\mathbf{g}\left(  \mathbf{x}_{t}\right)  }\frac{\partial}{\partial
x_{t}^{\beta}}\det\mathbf{g}\left(  \mathbf{x}_{t}\right)  -\frac{1}%
{2}DA^{\alpha}\left(  \mathbf{x}_{t}\right)  +\sqrt{2D}q_{a}^{\alpha}\left(
\mathbf{x}_{t}\right)  \mathbf{\xi}_{t}^{a}%
\end{equation}
where as usual $q_{a}^{\alpha}\left(  \mathbf{x}_{t}\right)  q_{a}^{\beta
}\left(  \mathbf{x}_{t}\right)  =g^{\alpha\beta}\left(  \mathbf{x}_{t}\right)
.$The Ito Fokker-Planck equation and sde are, respectively,%
\begin{equation}
\frac{\partial}{\partial t}P_{t}\left(  \mathbf{x}\right)  =\frac{\partial
}{\partial x^{\alpha}}\left(  g^{\alpha\beta}\left(  \mathbf{x}\right)
D\frac{\partial\beta F\left(  \mathbf{x}\right)  }{\partial x^{\beta}}%
-\frac{1}{2}Dg^{\alpha\beta}\left(  \mathbf{x}\right)  \frac{1}{\det
\mathbf{g}\left(  \mathbf{x}\right)  }\frac{\partial}{\partial x^{\beta}}%
\det\mathbf{g}\left(  \mathbf{x}\right)  +\frac{1}{2}DA^{\alpha}\left(
\mathbf{x}\right)  -D\frac{\partial g^{\alpha\beta}\left(  \mathbf{x}\right)
}{\partial x^{\beta}}+D\frac{\partial}{\partial x^{\beta}}g^{\alpha\beta
}\left(  \mathbf{x}\right)  \right)  P_{t}\left(  \mathbf{x}\right)
\end{equation}
and%
\begin{equation}
\frac{dx_{t}^{\alpha}}{dt}=-Dg^{\alpha\beta}\left(  \mathbf{x}_{t}\right)
\frac{\partial\beta F\left(  \mathbf{x}_{t}\right)  }{\partial x_{t}^{\beta}%
}+\frac{1}{2}Dg^{\alpha\beta}\left(  \mathbf{x}_{t}\right)  \frac{1}%
{\det\mathbf{g}\left(  \mathbf{x}_{t}\right)  }\frac{\partial}{\partial
x_{t}^{\beta}}\det\mathbf{g}\left(  \mathbf{x}_{t}\right)  -\frac{1}%
{2}DA^{\alpha}\left(  \mathbf{x}_{t}\right)  +D\frac{\partial g^{\alpha\beta
}\left(  \mathbf{x}\right)  }{\partial x^{\beta}}+\sqrt{2D}q_{a}^{\alpha
}\left(  \mathbf{x}_{t}\right)  \mathbf{\xi}_{t}^{a}.
\end{equation}

\subsection{The noise auto-correlation}

The noise autocorrelation function is
\begin{equation}
\int q_{a}^{\alpha}\left(  \mathbf{r}\right)  \left(  \int q_{a}^{\beta
}\left(  \mathbf{r}^{\prime}\right)  \delta\left(  \mathbf{r}-\mathbf{r}%
^{\prime}\right)  d\mathbf{r}^{\prime}\right)  d\mathbf{r}%
\end{equation}
and it is understood that the operators act on everything to their right. Now,
using the fact that the operator $\left(  \mathbf{\nabla}\cdot\rho\left(
\mathbf{r;x}\right)  \mathbf{\nabla}\right)  ^{-1}$ is self adjoint, (i.e. for
any two test functions $f\left(  \mathbf{r}\right)  $ and $g\left(
\mathbf{r}\right)  $ vanishing on $\partial V$,%
\begin{equation}
\int_{V}f\left(  \mathbf{r}\right)  \left(  \mathbf{\nabla}\cdot\rho\left(
\mathbf{r;x}\right)  \mathbf{\nabla}\right)  ^{-1}g\left(  \mathbf{r}\right)
d\mathbf{r=}\int_{V}g\left(  \mathbf{r}\right)  \left(  \mathbf{\nabla}%
\cdot\rho\left(  \mathbf{r;x}\right)  \mathbf{\nabla}\right)  ^{-1}f\left(
\mathbf{r}\right)  d\mathbf{r,}%
\end{equation}
as shown below in Appendix \ref{self} ), one has that
\begin{align}
&  \int q_{a}^{\beta}\left(  \mathbf{r}^{\prime}\right)  \delta\left(
\mathbf{r}-\mathbf{r}^{\prime}\right)  d\mathbf{r}^{\prime}\\
&  \mathbf{=}\mathbf{-}\int g^{\beta\gamma}\left(  \mathbf{x}\right)
\frac{\partial\rho\left(  \mathbf{r}^{\prime}\mathbf{;x}\right)  }{\partial
x^{\gamma}}\left(  \mathbf{\nabla}^{\prime}\cdot\rho\left(  \mathbf{r}%
^{\prime}\mathbf{;x}\right)  \mathbf{\nabla}^{\prime}\right)  ^{-1}%
\frac{\partial}{\partial r_{a}^{\prime}}\sqrt{2D\rho\left(  \mathbf{r}%
^{\prime}\mathbf{;x}\right)  }\delta\left(  \mathbf{r}-\mathbf{r}^{\prime
}\right)  d\mathbf{r}^{\prime}\nonumber\\
&  \mathbf{=}\mathbf{-}g^{\beta\gamma}\left(  \mathbf{x}\right)  \int%
\delta\left(  \mathbf{r}-\mathbf{r}^{\prime}\right)  \sqrt{2D\rho\left(
\mathbf{r}^{\prime}\mathbf{;x}\right)  }\frac{\partial}{\partial r_{a}%
^{\prime}}\left(  \mathbf{\nabla}^{\prime}\cdot\rho\left(  \mathbf{r}^{\prime
}\mathbf{;x}\right)  \mathbf{\nabla}^{\prime}\right)  ^{-1}\frac{\partial
\rho\left(  \mathbf{r}^{\prime}\mathbf{;x}\right)  }{\partial x^{\gamma}%
}d\mathbf{r}^{\prime}\nonumber\\
&  \mathbf{=}\mathbf{-}g^{\beta\gamma}\left(  \mathbf{x}\right)  \sqrt
{2D\rho\left(  \mathbf{r;x}\right)  }\frac{\partial}{\partial r_{a}}\left(
\mathbf{\nabla}\cdot\rho\left(  \mathbf{r;x}\right)  \mathbf{\nabla}\right)
^{-1}\frac{\partial\rho\left(  \mathbf{r;x}\right)  }{\partial x^{\gamma}%
}\nonumber
\end{align}
So%
\begin{align}
&  \int_{V}\int_{V}q_{a}^{\alpha}\left(  \mathbf{r}\right)  q_{a}^{\beta
}\left(  \mathbf{r}^{\prime}\right)  d\mathbf{r}d\mathbf{r}^{\prime}\\
&  =\int_{V}\left(  g^{\alpha\gamma}\left(  \mathbf{x}\right)  \frac
{\partial\rho\left(  \mathbf{r;}x\right)  }{\partial x^{\gamma}}\left(
\mathbf{\nabla}\cdot\rho\left(  \mathbf{r;x}\right)  \mathbf{\nabla}\right)
^{-1}\frac{\partial}{\partial r_{a}}\sqrt{2D\rho\left(  \mathbf{r;x}\right)
}\right) \nonumber\\
&  \times\left(  g^{\beta\sigma}\left(  \mathbf{x}\right)  \sqrt{2D\rho\left(
\mathbf{r;x}\right)  }\frac{\partial}{\partial r_{a}}\left(  \mathbf{\nabla
}\cdot\rho\left(  \mathbf{r;x}\right)  \mathbf{\nabla}\right)  ^{-1}%
\frac{\partial\rho\left(  \mathbf{r;}x\right)  }{\partial x^{\sigma}}\right)
d\mathbf{r}\nonumber\\
&  =2Dg^{\alpha\gamma}\left(  \mathbf{x}\right)  g^{\beta\sigma}\left(
\mathbf{x}\right)  \int_{V}\frac{\partial\rho\left(  \mathbf{r;}x\right)
}{\partial x^{\gamma}}\left(  \mathbf{\nabla}\cdot\rho\left(  \mathbf{r;x}%
\right)  \mathbf{\nabla}\right)  ^{-1}\left(  \mathbf{\nabla}\cdot\rho\left(
\mathbf{r;x}\right)  \mathbf{\nabla}\right)  \left(  \mathbf{\nabla}\cdot
\rho\left(  \mathbf{r;x}\right)  \mathbf{\nabla}\right)  ^{-1}\frac
{\partial\rho\left(  \mathbf{r;}x\right)  }{\partial x^{\sigma}}%
d\mathbf{r}\nonumber\\
&  =2Dg^{\alpha\gamma}\left(  \mathbf{x}\right)  g^{\beta\sigma}\left(
\mathbf{x}\right)  \int_{V}\frac{\partial\rho\left(  \mathbf{r;}x\right)
}{\partial x^{\gamma}}\left(  \mathbf{\nabla}\cdot\rho\left(  \mathbf{r;x}%
\right)  \mathbf{\nabla}\right)  ^{-1}\frac{\partial\rho\left(  \mathbf{r;}%
x\right)  }{\partial x^{\sigma}}d\mathbf{r}\nonumber\\
&  =2Dg^{\alpha\gamma}\left(  \mathbf{x}\right)  g^{\beta\sigma}\left(
\mathbf{x}\right)  g_{\gamma\sigma}\left(  \mathbf{x}\right) \nonumber\\
&  =2Dg^{\alpha\beta}\left(  \mathbf{x}\right) \nonumber
\end{align}

We also need%
\begin{align}
&  \int_{V}d\mathbf{r\;}q_{a}^{\alpha}\left(  \mathbf{r}\right)  \left(
\frac{\partial}{\partial x^{\beta}}\int d\mathbf{r}^{\prime}\;q_{a}^{\beta
}\left(  \mathbf{r}^{\prime}\right)  \delta\left(  \mathbf{r}-\mathbf{r}%
^{\prime}\right)  \right) \\
&  =\mathbf{-}\int_{V}d\mathbf{r\;}q_{a}^{\alpha}\left(  \mathbf{r}\right)
\frac{\partial g^{\beta\gamma}}{\partial x^{\beta}}\sqrt{2D\rho\left(
\mathbf{r;x}\right)  }\frac{\partial}{\partial r_{a}}\left(  \mathbf{\nabla
}\cdot\rho\left(  \mathbf{r;x}\right)  \mathbf{\nabla}\right)  ^{-1}%
\frac{\partial\rho\left(  \mathbf{r;}x\right)  }{\partial x^{\gamma}%
}\nonumber\\
&  \mathbf{-}\int_{V}d\mathbf{r\;}q_{a}^{\alpha}\left(  \mathbf{r}\right)
g^{\beta\gamma}\left(  \frac{\partial}{\partial x^{\beta}}\sqrt{2D\rho\left(
\mathbf{r;x}\right)  }\right)  \frac{\partial}{\partial r_{a}}\left(
\mathbf{\nabla}\cdot\rho\left(  \mathbf{r;x}\right)  \mathbf{\nabla}\right)
^{-1}\frac{\partial\rho\left(  \mathbf{r;}x\right)  }{\partial x^{\gamma}%
}\nonumber\\
&  \mathbf{-}\int_{V}d\mathbf{r\;}q_{a}^{\alpha}\left(  \mathbf{r}\right)
g^{\beta\gamma}\sqrt{2D\rho\left(  \mathbf{r;x}\right)  }\frac{\partial
}{\partial r_{a}}\frac{\partial}{\partial x^{\beta}}\left(  \mathbf{\nabla
}\cdot\rho\left(  \mathbf{r;x}\right)  \mathbf{\nabla}\right)  ^{-1}%
\frac{\partial\rho\left(  \mathbf{r;}x\right)  }{\partial x^{\gamma}}\nonumber
\end{align}
The first two terms follow directly from the previous calculation giving%
\begin{align}
&  \int_{V}d\mathbf{r\;}q_{a}^{\alpha}\left(  \mathbf{r}\right)  \left(
\frac{\partial}{\partial x^{\beta}}\int_{V}d\mathbf{r}^{\prime}\;q_{a}^{\beta
}\left(  \mathbf{r}^{\prime}\right)  \delta\left(  \mathbf{r}-\mathbf{r}%
^{\prime}\right)  \right) \\
&  =2Dg^{\alpha\gamma}\left(  \mathbf{x}\right)  \frac{\partial g^{\beta
\sigma}\left(  \mathbf{x}\right)  }{\partial x^{\beta}}g_{\gamma\sigma}\left(
\mathbf{x}\right) \nonumber\\
&  \mathbf{-}Dg^{\beta\sigma}\left(  \mathbf{x}\right)  g^{\alpha\gamma
}\left(  \mathbf{x}\right)  \int_{V}d\mathbf{r\;}\frac{\partial\rho\left(
\mathbf{r;}x\right)  }{\partial x^{\gamma}}\left(  \mathbf{\nabla}\cdot
\rho\left(  \mathbf{r;x}\right)  \mathbf{\nabla}\right)  ^{-1}\left(
\mathbf{\nabla}\cdot\frac{\partial\rho\left(  \mathbf{r;x}\right)  }{\partial
x^{\beta}}\mathbf{\nabla}\right)  \left(  \mathbf{\nabla}\cdot\rho\left(
\mathbf{r;x}\right)  \mathbf{\nabla}\right)  ^{-1}\frac{\partial\rho\left(
\mathbf{r;}x\right)  }{\partial x^{\sigma}}\nonumber\\
&  \mathbf{-}2Dg^{\alpha\gamma}\left(  \mathbf{x}\right)  g^{\beta\sigma
}\left(  \mathbf{x}\right)  \int_{V}d\mathbf{r\;}\frac{\partial\rho\left(
\mathbf{r;}x\right)  }{\partial x^{\gamma}}\frac{\partial}{\partial x^{\beta}%
}\left(  \mathbf{\nabla}\cdot\rho\left(  \mathbf{r;x}\right)  \mathbf{\nabla
}\right)  ^{-1}\frac{\partial\rho\left(  \mathbf{r;}x\right)  }{\partial
x^{\sigma}}\nonumber
\end{align}
or%
\begin{align}
&  \int_{V}d\mathbf{r\;}q_{a}^{\alpha}\left(  \mathbf{r}\right)  \left(
\frac{\partial}{\partial x^{\beta}}\int_{V}d\mathbf{r}^{\prime}\;q_{a}^{\beta
}\left(  \mathbf{r}^{\prime}\right)  \delta\left(  \mathbf{r}-\mathbf{r}%
^{\prime}\right)  \right) \\
&  =2Dg^{\alpha\gamma}\left(  \mathbf{x}\right)  \frac{\partial g^{\beta
\sigma}\left(  \mathbf{x}\right)  }{\partial x^{\beta}}g_{\gamma\sigma}\left(
\mathbf{x}\right) \nonumber\\
&  \mathbf{-}Dg^{\beta\sigma}\left(  \mathbf{x}\right)  g^{\alpha\gamma
}\left(  \mathbf{x}\right)  \int_{V}d\mathbf{r\;}\frac{\partial\rho\left(
\mathbf{r;x}\right)  }{\partial x^{\gamma}}\left(  \mathbf{\nabla}\cdot
\rho\left(  \mathbf{r;x}\right)  \mathbf{\nabla}\right)  ^{-1}\left(
\mathbf{\nabla}\cdot\frac{\partial\rho\left(  \mathbf{r;x}\right)  }{\partial
x^{\beta}}\mathbf{\nabla}\right)  \left(  \mathbf{\nabla}\cdot\rho\left(
\mathbf{r;x}\right)  \mathbf{\nabla}\right)  ^{-1}\frac{\partial\rho\left(
\mathbf{r;x}\right)  }{\partial x^{\sigma}}\nonumber\\
&  \mathbf{-}2Dg^{\alpha\gamma}\left(  \mathbf{x}\right)  g^{\beta\sigma
}\left(  \mathbf{x}\right)  \frac{\partial}{\partial x^{\beta}}g_{\gamma
\sigma}\left(  \mathbf{x}\right) \nonumber\\
&  \mathbf{+}2Dg^{\alpha\gamma}\left(  \mathbf{x}\right)  g^{\beta\sigma
}\left(  \mathbf{x}\right)  \int_{V}d\mathbf{r\;}\frac{\partial^{2}\rho\left(
\mathbf{r;x}\right)  }{\partial x^{\beta}\partial x^{\gamma}}\left(
\mathbf{\nabla}\cdot\rho\left(  \mathbf{r;x}\right)  \mathbf{\nabla}\right)
^{-1}\frac{\partial\rho\left(  \mathbf{r;x}\right)  }{\partial x^{\sigma}%
}.\nonumber
\end{align}
Thus%
\begin{align}
&  \int_{V}d\mathbf{r\;}q_{a}^{\alpha}\left(  \mathbf{r}\right)  \left(
\frac{\partial}{\partial x^{\beta}}\int_{V}d\mathbf{r}^{\prime}\;q_{a}^{\beta
}\left(  \mathbf{r}^{\prime}\right)  \delta\left(  \mathbf{r}-\mathbf{r}%
^{\prime}\right)  \right) \\
&  =\mathbf{-}Dg^{\beta\sigma}\left(  \mathbf{x}\right)  g^{\alpha\gamma
}\left(  \mathbf{x}\right)  \int_{V}d\mathbf{r\;}\frac{\partial\rho\left(
\mathbf{r;}x\right)  }{\partial x^{\gamma}}\left(  \mathbf{\nabla}\cdot
\rho\left(  \mathbf{r;x}\right)  \mathbf{\nabla}\right)  ^{-1}\left(
\mathbf{\nabla}\cdot\frac{\partial\rho\left(  \mathbf{r;x}\right)  }{\partial
x^{\beta}}\mathbf{\nabla}\right)  \left(  \mathbf{\nabla}\cdot\rho\left(
\mathbf{r;x}\right)  \mathbf{\nabla}\right)  ^{-1}\frac{\partial\rho\left(
\mathbf{r;}x\right)  }{\partial x^{\sigma}}\nonumber\\
&  \mathbf{+}2Dg^{\alpha\gamma}\left(  \mathbf{x}\right)  g^{\beta\sigma
}\left(  \mathbf{x}\right)  \int_{V}d\mathbf{r\;}\frac{\partial^{2}\rho\left(
\mathbf{r;}x\right)  }{\partial x^{\beta}\partial x^{\gamma}}\left(
\mathbf{\nabla}\cdot\rho\left(  \mathbf{r;x}\right)  \mathbf{\nabla}\right)
^{-1}\frac{\partial\rho\left(  \mathbf{r;}x\right)  }{\partial x^{\sigma}%
}\nonumber
\end{align}
Using the symmetry of the metric,
\begin{align}
&  2Dg^{\alpha\gamma}\left(  \mathbf{x}\right)  g^{\beta\sigma}\left(
\mathbf{x}\right)  \int_{V}d\mathbf{r\;}\frac{\partial^{2}\rho\left(
\mathbf{r;x}\right)  }{\partial x^{\beta}\partial x^{\gamma}}\left(
\mathbf{\nabla}\cdot\rho\left(  \mathbf{r;x}\right)  \mathbf{\nabla}\right)
^{-1}\frac{\partial\rho\left(  \mathbf{r;x}\right)  }{\partial x^{\sigma}}\\
&  =Dg^{\alpha\gamma}\left(  \mathbf{x}\right)  g^{\beta\sigma}\left(
\mathbf{x}\right)  \int_{V}d\mathbf{r\;}\frac{\partial^{2}\rho\left(
\mathbf{r;x}\right)  }{\partial x^{\beta}\partial x^{\gamma}}\left(
\mathbf{\nabla}\cdot\rho\left(  \mathbf{r;x}\right)  \mathbf{\nabla}\right)
^{-1}\frac{\partial\rho\left(  \mathbf{r;x}\right)  }{\partial x^{\sigma}%
}\nonumber\\
&  \mathbf{+}Dg^{\alpha\gamma}\left(  \mathbf{x}\right)  g^{\beta\sigma
}\left(  \mathbf{x}\right)  \int_{V}d\mathbf{r\;}\frac{\partial^{2}\rho\left(
\mathbf{r;x}\right)  }{\partial x^{\sigma}\partial x^{\gamma}}\left(
\mathbf{\nabla}\cdot\rho\left(  \mathbf{r;x}\right)  \mathbf{\nabla}\right)
^{-1}\frac{\partial\rho\left(  \mathbf{r;x}\right)  }{\partial x^{\beta}%
}\nonumber\\
&  =Dg^{\alpha\gamma}\left(  \mathbf{x}\right)  g^{\beta\sigma}\left(
\mathbf{x}\right)  \int_{V}d\mathbf{r\;}\frac{\partial^{2}\rho\left(
\mathbf{r;x}\right)  }{\partial x^{\beta}\partial x^{\gamma}}\left(
\mathbf{\nabla}\cdot\rho\left(  \mathbf{r;x}\right)  \mathbf{\nabla}\right)
^{-1}\frac{\partial\rho\left(  \mathbf{r;x}\right)  }{\partial x^{\sigma}%
}\nonumber\\
&  \mathbf{+}Dg^{\alpha\gamma}\left(  \mathbf{x}\right)  g^{\beta\sigma
}\left(  \mathbf{x}\right)  \int_{V}d\mathbf{r\;}\frac{\partial\rho\left(
\mathbf{r;x}\right)  }{\partial x^{\beta}}\left(  \mathbf{\nabla}\cdot
\rho\left(  \mathbf{r;x}\right)  \mathbf{\nabla}\right)  ^{-1}\frac
{\partial^{2}\rho\left(  \mathbf{r;x}\right)  }{\partial x^{\sigma}\partial
x^{\gamma}}\nonumber\\
&  =Dg^{\alpha\gamma}\left(  \mathbf{x}\right)  g^{\beta\sigma}\left(
\mathbf{x}\right)  \frac{\partial}{\partial x^{\gamma}}\int_{V}d\mathbf{r\;}%
\frac{\partial\rho\left(  \mathbf{r;x}\right)  }{\partial x^{\beta}}\left(
\mathbf{\nabla}\cdot\rho\left(  \mathbf{r;x}\right)  \mathbf{\nabla}\right)
^{-1}\frac{\partial\rho\left(  \mathbf{r;x}\right)  }{\partial x^{\sigma}%
}\nonumber\\
&  -Dg^{\alpha\gamma}\left(  \mathbf{x}\right)  g^{\beta\sigma}\left(
\mathbf{x}\right)  \int_{V}d\mathbf{r\;}\frac{\partial\rho\left(
\mathbf{r;x}\right)  }{\partial x^{\beta}}\left[  \frac{\partial}{\partial
x^{\gamma}}\left(  \mathbf{\nabla}\cdot\rho\left(  \mathbf{r;x}\right)
\mathbf{\nabla}\right)  ^{-1}\right]  \frac{\partial\rho\left(  \mathbf{r;x}%
\right)  }{\partial x^{\sigma}}\nonumber
\end{align}
or%
\begin{align}
&  =-Dg^{\alpha\gamma}\left(  \mathbf{x}\right)  g^{\beta\sigma}\left(
\mathbf{x}\right)  \frac{\partial}{\partial x^{\gamma}}g_{\beta\sigma}\left(
\mathbf{x}\right) \\
&  +Dg^{\alpha\gamma}\left(  \mathbf{x}\right)  g^{\beta\sigma}\left(
\mathbf{x}\right)  \int_{V}d\mathbf{r\;}\frac{\partial\rho\left(
\mathbf{r;x}\right)  }{\partial x^{\beta}}\left[  \left(  \mathbf{\nabla}%
\cdot\rho\left(  \mathbf{r;x}_{t}\right)  \mathbf{\nabla}\right)  ^{-1}\left(
\mathbf{\nabla}\cdot\frac{\partial\rho\left(  \mathbf{r;x}_{t}\right)
}{\partial x^{\gamma}}\mathbf{\nabla}\right)  \left(  \mathbf{\nabla}\cdot
\rho\left(  \mathbf{r;x}\right)  \mathbf{\nabla}\right)  ^{-1}\right]
\frac{\partial\rho\left(  \mathbf{r;}x\right)  }{\partial x^{\sigma}}\nonumber
\end{align}
Putting these pieces together%
\begin{align}
&  \int_{V}d\mathbf{r\;}q_{a}^{\alpha}\left(  \mathbf{r}\right)  \left(
\frac{\partial}{\partial x^{\beta}}\int d\mathbf{r}^{\prime}\;q_{a}^{\beta
}\left(  \mathbf{r}^{\prime}\right)  \delta\left(  \mathbf{r}-\mathbf{r}%
^{\prime}\right)  \right) \\
&  =\mathbf{-}Dg^{\beta\sigma}\left(  \mathbf{x}\right)  g^{\alpha\gamma
}\left(  \mathbf{x}\right)  \int_{V}d\mathbf{r\;}\frac{\partial\rho\left(
\mathbf{r;}x\right)  }{\partial x^{\gamma}}\left(  \mathbf{\nabla}\cdot
\rho\left(  \mathbf{r;x}\right)  \mathbf{\nabla}\right)  ^{-1}\left(
\mathbf{\nabla}\cdot\frac{\partial\rho\left(  \mathbf{r;x}\right)  }{\partial
x^{\beta}}\mathbf{\nabla}\right)  \left(  \mathbf{\nabla}\cdot\rho\left(
\mathbf{r;x}\right)  \mathbf{\nabla}\right)  ^{-1}\frac{\partial\rho\left(
\mathbf{r;}x\right)  }{\partial x^{\sigma}}\nonumber\\
&  -Dg^{\alpha\gamma}\left(  \mathbf{x}\right)  g^{\beta\sigma}\left(
\mathbf{x}\right)  \frac{\partial}{\partial x^{\gamma}}g_{\beta\sigma}\left(
\mathbf{x}\right) \nonumber\\
&  +Dg^{\alpha\gamma}\left(  \mathbf{x}\right)  g^{\beta\sigma}\left(
\mathbf{x}\right)  \int_{V}d\mathbf{r\;}\frac{\partial\rho\left(
\mathbf{r;}x\right)  }{\partial x^{\beta}}\left[  \left(  \mathbf{\nabla}%
\cdot\rho\left(  \mathbf{r;x}\right)  \mathbf{\nabla}\right)  ^{-1}\left(
\mathbf{\nabla}\cdot\frac{\partial\rho\left(  \mathbf{r;x}\right)  }{\partial
x^{\gamma}}\mathbf{\nabla}\right)  \left(  \mathbf{\nabla}\cdot\rho\left(
\mathbf{r;x}\right)  \mathbf{\nabla}\right)  ^{-1}\right]  \frac{\partial
\rho\left(  \mathbf{r;}x\right)  }{\partial x^{\sigma}}\nonumber
\end{align}
or%
\begin{align}
&  \int_{V}d\mathbf{r\;}q_{a}^{\alpha}\left(  \mathbf{r}\right)  \left(
\frac{\partial}{\partial x^{\beta}}\int_{V}d\mathbf{r}^{\prime}\;q_{a}^{\beta
}\left(  \mathbf{r}^{\prime}\right)  \delta\left(  \mathbf{r}-\mathbf{r}%
^{\prime}\right)  \right) \\
&  =-Dg^{\alpha\gamma}\left(  \mathbf{x}\right)  g^{\beta\sigma}\left(
\mathbf{x}\right)  \frac{\partial}{\partial x^{\gamma}}g_{\beta\sigma}\left(
\mathbf{x}\right) \nonumber\\
&  -D\left(  g^{\alpha\beta}\left(  \mathbf{x}\right)  g^{\gamma\sigma}\left(
\mathbf{x}\right)  -g^{\alpha\gamma}\left(  \mathbf{x}\right)  g^{\beta\sigma
}\left(  \mathbf{x}\right)  \right) \nonumber\\
&  \times\int_{V}d\mathbf{r\;}\frac{\partial\rho\left(  \mathbf{r;x}\right)
}{\partial x^{\gamma}}\left(  \mathbf{\nabla}\cdot\rho\left(  \mathbf{r;x}%
\right)  \mathbf{\nabla}\right)  ^{-1}\left(  \mathbf{\nabla}\cdot
\frac{\partial\rho\left(  \mathbf{r;x}\right)  }{\partial x^{\beta}%
}\mathbf{\nabla}\right)  \left(  \mathbf{\nabla}\cdot\rho\left(
\mathbf{r;x}\right)  \mathbf{\nabla}\right)  ^{-1}\frac{\partial\rho\left(
\mathbf{r;x}\right)  }{\partial x^{\sigma}}\nonumber
\end{align}
More symmetrically,%
\begin{align}
&  \int_{V}d\mathbf{r\;}q_{a}^{\alpha}\left(  \mathbf{r}\right)  \left(
\frac{\partial}{\partial x^{\beta}}\int_{V}d\mathbf{r}^{\prime}\;q_{a}^{\beta
}\left(  \mathbf{r}^{\prime}\right)  \delta\left(  \mathbf{r}-\mathbf{r}%
^{\prime}\right)  \right) \\
&  =-Dg^{\alpha\gamma}\left(  \mathbf{x}\right)  g^{\beta\sigma}\left(
\mathbf{x}\right)  \frac{\partial}{\partial x^{\gamma}}g_{\beta\sigma}\left(
\mathbf{x}\right) \nonumber\\
&  +D\left(  g^{\alpha\beta}\left(  \mathbf{x}\right)  g^{\gamma\sigma}\left(
\mathbf{x}\right)  -g^{\alpha\gamma}\left(  \mathbf{x}\right)  g^{\beta\sigma
}\left(  \mathbf{x}\right)  \right) \nonumber\\
&  \times\int_{V}d\mathbf{r\;}\frac{\partial\rho\left(  \mathbf{r;x}\right)
}{\partial x^{\beta}}\left(  \mathbf{\nabla}\left[  \left(  \mathbf{\nabla
}\cdot\rho\left(  \mathbf{r;x}\right)  \mathbf{\nabla}\right)  ^{-1}%
\frac{\partial\rho\left(  \mathbf{r;x}\right)  }{\partial x_{t}^{\gamma}%
}\right]  \right)  \cdot\left(  \mathbf{\nabla}\left[  \left(  \mathbf{\nabla
}\cdot\rho\left(  \mathbf{r;x}\right)  \mathbf{\nabla}\right)  ^{-1}%
\frac{\partial\rho\left(  \mathbf{r;x}\right)  }{\partial x_{t}^{\sigma}%
}\right]  \right) \nonumber
\end{align}
or finally%
\begin{align}
&  \int_{V}d\mathbf{r\;}q_{a}^{\alpha}\left(  \mathbf{r}\right)  \left(
\frac{\partial}{\partial x^{\beta}}\int d\mathbf{r}^{\prime}\;q_{a}^{\beta
}\left(  \mathbf{r}^{\prime}\right)  \delta\left(  \mathbf{r}-\mathbf{r}%
^{\prime}\right)  \right) \\
&  =-Dg^{\alpha\gamma}\left(  \mathbf{x}\right)  \frac{1}{\det\mathbf{g}%
\left(  \mathbf{x}\right)  }\frac{\partial}{\partial x^{\gamma}}\det
\mathbf{g}\left(  \mathbf{x}\right) \nonumber\\
&  +D\left(  g^{\alpha\beta}\left(  \mathbf{x}\right)  g^{\gamma\sigma}\left(
\mathbf{x}\right)  -g^{\alpha\gamma}\left(  \mathbf{x}\right)  g^{\beta\sigma
}\left(  \mathbf{x}\right)  \right) \nonumber\\
&  \times\int_{V}d\mathbf{r\;}\frac{\partial\rho\left(  \mathbf{r;x}\right)
}{\partial x^{\beta}}\left(  \mathbf{\nabla}\left[  \left(  \mathbf{\nabla
}\cdot\rho\left(  \mathbf{r;x}\right)  \mathbf{\nabla}\right)  ^{-1}%
\frac{\partial\rho\left(  \mathbf{r;x}\right)  }{\partial x_{t}^{\gamma}%
}\right]  \right)  \cdot\left(  \mathbf{\nabla}\left[  \left(  \mathbf{\nabla
}\cdot\rho\left(  \mathbf{r;x}\right)  \mathbf{\nabla}\right)  ^{-1}%
\frac{\partial\rho\left(  \mathbf{r;x}\right)  }{\partial x_{t}^{\sigma}%
}\right]  \right) \nonumber
\end{align}

\subsection{Proof of self-adjointness}

\label{self} To prove that the inverse operator is self-adjoint, consider two
arbitrary test functions $f\left(  \mathbf{r}\right)  $ and $g\left(
\mathbf{r}\right)  $ ,%
\begin{align}
\int_{V}f\left(  \mathbf{r}\right)  \left(  \mathbf{\nabla}\cdot\rho\left(
\mathbf{r;x}\right)  \mathbf{\nabla}\right)  ^{-1}g\left(  \mathbf{r}\right)
d\mathbf{r}  &  =\int_{V}\left[  \left(  \mathbf{\nabla}\cdot\rho\left(
\mathbf{r;x}\right)  \mathbf{\nabla}\right)  \left(  \mathbf{\nabla}\cdot
\rho\left(  \mathbf{r;x}\right)  \mathbf{\nabla}\right)  ^{-1}f\left(
\mathbf{r}\right)  \right]  \left(  \mathbf{\nabla}\cdot\rho\left(
\mathbf{r;x}\right)  \mathbf{\nabla}\right)  ^{-1}g\left(  \mathbf{r}\right)
d\mathbf{r}\\
&  =-\int_{V}\left[  \rho\left(  \mathbf{r;x}\right)  \mathbf{\nabla}\left(
\mathbf{\nabla}\cdot\rho\left(  \mathbf{r;x}\right)  \mathbf{\nabla}\right)
^{-1}f\left(  \mathbf{r}\right)  \right]  \cdot\mathbf{\nabla}\left(
\mathbf{\nabla}\cdot\rho\left(  \mathbf{r;x}\right)  \mathbf{\nabla}\right)
^{-1}g\left(  \mathbf{r}\right)  d\mathbf{r}\nonumber
\end{align}
provided the surface term vanishes,%
\begin{equation}
0=\int_{\partial V}\rho\left(  \mathbf{r;x}\right)  \left[  \mathbf{\nabla
}\left(  \mathbf{\nabla}\cdot\rho\left(  \mathbf{r;x}\right)  \mathbf{\nabla
}\right)  ^{-1}f\left(  \mathbf{r}\right)  \right]  \left(  \mathbf{\nabla
}\cdot\rho\left(  \mathbf{r;x}\right)  \mathbf{\nabla}\right)  ^{-1}g\left(
\mathbf{r}\right)  \cdot d\mathbf{S}%
\end{equation}
Then a second integration by parts gives%
\begin{align}
\int_{V}f\left(  \mathbf{r}\right)  \left(  \mathbf{\nabla}\cdot\rho\left(
\mathbf{r;x}\right)  \mathbf{\nabla}\right)  ^{-1}g\left(  \mathbf{r}\right)
d\mathbf{r}  &  =-\int_{V}\left[  \rho\left(  \mathbf{r;x}\right)
\mathbf{\nabla}\left(  \mathbf{\nabla}\cdot\rho\left(  \mathbf{r;x}\right)
\mathbf{\nabla}\right)  ^{-1}f\left(  \mathbf{r}\right)  \right]
\cdot\mathbf{\nabla}\left(  \mathbf{\nabla}\cdot\rho\left(  \mathbf{r;x}%
\right)  \mathbf{\nabla}\right)  ^{-1}g\left(  \mathbf{r}\right)
d\mathbf{r}\\
&  =\int_{V}\left[  \left(  \mathbf{\nabla}\cdot\rho\left(  \mathbf{r;x}%
\right)  \mathbf{\nabla}\right)  ^{-1}f\left(  \mathbf{r}\right)  \right]
\left(  \mathbf{\nabla}\cdot\rho\left(  \mathbf{r;x}\right)  \mathbf{\nabla
}\right)  \left(  \mathbf{\nabla}\cdot\rho\left(  \mathbf{r;x}\right)
\mathbf{\nabla}\right)  ^{-1}g\left(  \mathbf{r}\right)  d\mathbf{r}%
\nonumber\\
&  =\int_{V}\left[  \left(  \mathbf{\nabla}\cdot\rho\left(  \mathbf{r;x}%
\right)  \mathbf{\nabla}\right)  ^{-1}f\left(  \mathbf{r}\right)  \right]
g\left(  \mathbf{r}\right)  d\mathbf{r}\nonumber
\end{align}
which is the desired result assuming another boundary term vanishes,%
\begin{equation}
0=-\int_{\partial V}\left[  \left(  \mathbf{\nabla}\cdot\rho\left(
\mathbf{r;x}\right)  \mathbf{\nabla}\right)  ^{-1}f\left(  \mathbf{r}\right)
\right]  \rho\left(  \mathbf{r;x}\right)  \mathbf{\nabla}\left(
\mathbf{\nabla}\cdot\rho\left(  \mathbf{r;x}\right)  \mathbf{\nabla}\right)
^{-1}g\left(  \mathbf{r}\right)  \cdot d\mathbf{S.}%
\end{equation}
The vanishing of both boundary terms follows froms the no-flux boundary
condition which can be formulated as follows. Define%
\begin{align}
\left(  \mathbf{\nabla}\cdot\rho\left(  \mathbf{r;x}\right)  \mathbf{\nabla
}\right)  ^{-1}f\left(  \mathbf{r}\right)   &  =\phi_{f}\left(  \mathbf{r}%
\right) \\
\left(  \mathbf{\nabla}\cdot\rho\left(  \mathbf{r;x}\right)  \mathbf{\nabla
}\right)  ^{-1}g\left(  \mathbf{r}\right)   &  =\phi_{g}\left(  \mathbf{r}%
\right) \nonumber
\end{align}
so that%
\begin{align}
\left(  \mathbf{\nabla}\cdot\rho\left(  \mathbf{r;x}\right)  \mathbf{\nabla
}\right)  \phi_{f}\left(  \mathbf{r}\right)   &  =f\left(  \mathbf{r}\right)
\\
\left(  \mathbf{\nabla}\cdot\rho\left(  \mathbf{r;x}\right)  \mathbf{\nabla
}\right)  \phi_{g}\left(  \mathbf{r}\right)   &  =g\left(  \mathbf{r}\right)
\nonumber
\end{align}
and the boundary terms can be written as%
\begin{align}
&  \int_{\partial V}\rho\left(  \mathbf{r;x}\right)  \phi_{g}\left(
\mathbf{r}\right)  \left[  \mathbf{\nabla}\phi_{f}\left(  \mathbf{r}\right)
\right]  \cdot d\mathbf{S}\\
&  -\int_{\partial V}\phi_{f}\left(  \mathbf{r}\right)  \rho\left(
\mathbf{r;x}\right)  \mathbf{\nabla}\phi_{g}\left(  \mathbf{r}\right)  \cdot
d\mathbf{S}\nonumber
\end{align}
but the no-flux boundary condition says that
\begin{equation}
\left[  \mathbf{\nabla}\phi_{f}\left(  \mathbf{r}\right)  \right]  \cdot
d\mathbf{S=}\left[  \mathbf{\nabla}\phi_{g}\left(  \mathbf{r}\right)  \right]
\cdot d\mathbf{S=0}%
\end{equation}
on the surface.

\end{document}